\documentclass[11pt]{article}
\usepackage[a4paper, width = 150mm, top = 30mm, bottom = 30mm]{geometry}
\usepackage{jheppub}
\usepackage{mathtools}
\usepackage{mathrsfs}
\usepackage{bbold}
\usepackage{float}
\usepackage{pgfplots}
\pgfplotsset{compat=1.18}
\usetikzlibrary{hobby}
\usetikzlibrary{decorations.markings}
\usepackage{makecell}
\usepackage{circuitikz}
\usepackage{subcaption}
\usepackage{ifthen}
\usepackage{braket}
\usepackage[bbgreekl]{mathbbol}

\newcommand{\bx}{\mathbf{x}}
\newcommand{\bk}{\mathbf{k}}

\newcommand{\A}{\scalebox{.9}{$\scriptscriptstyle A$}}
\newcommand{\B}{\scalebox{.9}{$\scriptscriptstyle B$}}

\newcommand{\Psmall}{\scalebox{.9}{$\scriptscriptstyle P$}}
\newcommand{\F}{\scalebox{.9}{$\scriptscriptstyle F$}}
\newcommand{\Pb}{\bar{\Psmall}}
\newcommand{\Fb}{\bar{\F}}

\newcommand{\sR}{{}_{\text{R}}}
\newcommand{\sL}{{}_{\text{L}}}

\newcommand{\bbG}{\mathbb{G}}
\newcommand{\bbGA}{\mathbb{G}_{\text{adv}}}
\newcommand{\bbGR}{\mathbb{G}_{\text{ret}}}
\newcommand{\tbbGA}{\widetilde{\mathbb{G}}_{\text{adv}}}
\newcommand{\tbbGR}{\widetilde{\mathbb{G}}_{\text{ret}}}

\newcommand{\Gin}{G^{\text{in}}}
\newcommand{\Gout}{G^{\text{out}}}
\newcommand{\tGin}{\widetilde{G}^{\text{in}}}
\newcommand{\tGout}{\widetilde{G}^{\text{out}}}
\newcommand{\Kin}{K^{\text{in}}}

\newcommand{\Pplus}{\mathbb{P}_{+}}
\newcommand{\Pmin}{\mathbb{P}_{-}}
\newcommand{\G}{\Gamma}

\newcommand{\Psibar}{\overline{\Psi}}
\newcommand{\psibar}{\overline{\psi}}
\newcommand{\psiP}{\psi_{\bar{\Psmall}}}
\newcommand{\psiF}{\psi_{\bar{\F}}}
\newcommand{\psibP}{\overline{\psi}_{\bar{\Psmall}}}
\newcommand{\psibF}{\overline{\psi}_{\bar{\F}}}

\newcommand{\bbS}{\mathbb{S}}
\newcommand{\bbSA}{\mathbb{S}_{\text{adv}}}
\newcommand{\bbSR}{\mathbb{S}_{\text{ret}}}
\newcommand{\bbSDD}{\mathbb{S}_{{}_\text{DD}}}

\newcommand{\bbW}{\mathfrak{W}}
\newcommand{\frakW}{\mathbb{W}}

\newcommand{\Sbo}{\overline{S}_0}
\newcommand{\Sin}{S^{\text{in}}}
\newcommand{\Sout}{S^{\text{out}}}
\newcommand{\Sbin}{\overline{S}^{\text{in}}}
\newcommand{\Sbout}{\overline{S}^{\text{out}}}

\newcommand{\Kret}{\mathfrak{K}_{\rm ret}}

\newcommand{\eF}{e_{\bar{\F}}}
\newcommand{\eP}{e_{\bar{\Psmall}}}
\newcommand{\fF}{f_{\bar{\F}}}
\newcommand{\fP}{f_{\bar{\Psmall}}}

\newcommand{\bulkalpha}{\scalebox{1.2}{$\scriptscriptstyle \bbalpha$}}
\newcommand{\bulkbeta}{\scalebox{1.2}{$\scriptscriptstyle \bbbeta$}}

\newcommand{\z}{\zeta}
\newcommand{\w}{\omega}

\newcommand*\diff{\mathop{}\!\mathrm{d}}
\newcommand{\projp}{\mathbb{P}_{_+}}
\newcommand{\projm}{\mathbb{P}_{_-}}
\newcommand{\thetaSK}{\Theta_{{}^{\text{SK}}}}
\newcommand{\fd}{{}_\text{FD}}
\newcommand{\be}{{}_\text{BE}}
\newcommand{\bbT}{\mathbb{T}}

\newcommand{\red}[1]{\textcolor{black}{#1}}

\title{\boldmath Open EFT for Interacting Fermions from Holography}
\author{Godwin Martin, Shivam K. Sharma}
\affiliation{International Centre for Theoretical Sciences (ICTS-TIFR)\\ 
Tata Institute of Fundamental Research, Shivakote, Hesaraghatta Hobli, Bengaluru 560089, India.}
\emailAdd{godwin.martin@icts.res.in, shivam.sharma@icts.res.in}  
\abstract{ We initiate the study of an open EFT for finite-temperature holographic systems with interacting fermions. In particular, we do this for Yukawa interactions in the bulk using the real-time formalism (grSK geometry). From the bulk perspective, this study corresponds to Yukawa scattering against a black hole, incorporating the effects of Hawking radiation. We derive a field theory in the exterior of the black hole and thus develop a Witten diagrammatic understanding of the scattering processes. This allows the explicit evaluation of boundary Schwinger-Keldysh (SK) correlators to arbitrary order, at tree-level in the bulk. Here we present explicitly the SK generating functional up to four-point functions. Finally, we represent the correlators in a column-vector representation that manifests microscopic unitarity and thermality, with all the statistical factors neatly folded in.}


\begin{document}

\newcommand{\Gsemicap}[4]{
\draw[dashed,{|[scale=1.8]}-] (#1 +  #3,#2 + #4)--(#3,#4);
}

\newcommand{\Gdiodearrow}[4]{
\draw[dashed,-{Triangle[scale=1.8,open]}] (#1,#2)--(#1 +  #3 , #2 + #4);
}

\newcommand{\Ssemicap}[4]{
\draw[thick,{|[scale=1.8]}-] (#1 +  #3,#2 + #4)--(#3,#4);
}

\newcommand{\Sdiodearrow}[4]{
\draw[thick,-{Triangle[scale=1.8,open]}] (#1,#2)--(#1 +  #3 , #2 + #4);
}

\newcommand{\Gdiode}[4]{
	\draw[dashed,-{Triangle[scale=1.8,open]}] (#1,#2)--(#1/2 +  #3/2 , #2/2 + #4/2);
	\draw[dashed,{|[scale=1.8]}-] (#1/2 +  #3/2,#2/2 + #4/2)--(#3,#4);
}

\newcommand{\Sdiode}[4]{
	\draw[-{Triangle[scale=1.8,open]}] (#1,#2)--(#1/2 +  #3/2 , #2/2 + #4/2);
	\draw[{|[scale=1.8]}-] (#1/2 +  #3/2,#2/2 + #4/2)--(#3,#4);
}

\newcommand{\threeptPFYuk}[3]{
\begin{tikzpicture}[scale=1]
    \coordinate (z) at (0,-0.5);
    \draw[blue] (-1.4,1)--(1.4,1);
    \ifthenelse{#1 > 0}{\Sdiode{-1.2}{1}{0}{-0.5}}{\Sdiode{0}{-0.5}{-1.2}{1}};
    \ifthenelse{#2 > 0}{\Gdiode{0}{1}{0}{-0.5}}{\Gdiode{0}{-0.5}{0}{1}};
    \ifthenelse{#3 > 0}{\Sdiode{1.2}{1}{0}{-0.5}}{\Sdiode{0}{-0.5}{1.2}{1}};
    \node at (z) {$\bullet$};
\end{tikzpicture}
}

\newcommand{\fourptScalExPFYuk}[5]{
\begin{tikzpicture}[scale=1.4]
	\draw[blue] (-1.75,0)--(1.75,0);
	\ifthenelse{#1 > 0}{\Sdiode{-1.5}{0}{-1}{-1}}{\Sdiode{-1}{-1}{-1.5}{0}}
	\ifthenelse{#2 > 0}{\Sdiode{-0.5}{0}{-1}{-1}}{\Sdiode{-1}{-1}{-0.5}{0}}
	\ifthenelse{#3 > 0}{\Sdiode{0.5}{0}{1}{-1}}{\Sdiode{1}{-1}{0.5}{0}}
	\ifthenelse{#4 > 0}{\Sdiode{1.5}{0}{1}{-1}}{\Sdiode{1}{-1}{1.5}{0}}
	\ifthenelse{#5 > 0}{\Gdiode{-1}{-1}{1}{-1}}{\Gdiode{1}{-1}{-1}{-1}}
	\node at (-1,-1) {$\bullet$};
	\node at (1,-1) {$\bullet$};
\end{tikzpicture}
}

\newcommand{\fourptFermExPFYuk}[5]{
\begin{tikzpicture}[scale=1.4]
	\draw[blue] (-1.75,0)--(1.75,0);
	\ifthenelse{#1 > 0}{\Gdiode{-1.5}{0}{-1}{-1}}{\Gdiode{-1}{-1}{-1.5}{0}}
	\ifthenelse{#2 > 0}{\Sdiode{-0.5}{0}{-1}{-1}}{\Sdiode{-1}{-1}{-0.5}{0}}
	\ifthenelse{#3 > 0}{\Gdiode{0.5}{0}{1}{-1}}{\Gdiode{1}{-1}{0.5}{0}}
	\ifthenelse{#4 > 0}{\Sdiode{1.5}{0}{1}{-1}}{\Sdiode{1}{-1}{1.5}{0}}
	\ifthenelse{#5 > 0}{\Sdiode{-1}{-1}{1}{-1}}{\Sdiode{1}{-1}{-1}{-1}}
	\node at (-1,-1) {$\bullet$};
	\node at (1,-1) {$\bullet$};
\end{tikzpicture}
}

\maketitle

\section{Introduction}\label{sec:introduction}

Understanding phases of matter without a quasiparticle description is a challenging problem \cite{varma2002singular, Lee:2017njh}, spanning from the high-$T_c$ superconductors in condensed matter to the \textit{quark-gluon plasma} (QGP) in high-energy physics. This is largely due to strong-coupling physics which makes traditional perturbative techniques ineffective. The AdS/CFT correspondence \cite{Maldacena:1997re, Gubser:1998bc, Witten:1998qj} provides a lifeline with solvable toy models \cite{Hartnoll:2016apf, Blake:2022uyo}, offering a perturbative description through gravitational duals. In particular, it allows us to study strongly coupled systems through classical computations in black hole geometries. In fact, thermal states in these holographic systems correspond to black holes in gravity. 

Recent developments employ effective field theory (EFT) techniques \cite{Polchinski:1992ed, Polchinski:1993ii} in such holographic models \cite{Lee:2008xf,Lee:2009epi, Liu:2009dm, Faulkner:2009wj, Iqbal:2011ae}, focusing on the dynamics of probes (e.g. electrons). This requires adopting the perspective of open effective field theories (open EFTs). The approach involves modelling the \emph{bath} as a holographic system and deriving an EFT for the \emph{probe} coupled to this bath. Holographic baths are strongly interacting and also maximally \emph{forgetful}, they naturally give rise to a local EFT for the probe. In contrast, weakly coupled baths have \emph{memory} and result in a non-local EFT. This highlights the role played by holography in formulating local open EFTs. The aim here is to extract the influence of the holographic system on the probe, captured by the \textit{Influence functional} \cite{Feynman:1963fq}. This functional is commonly derived using the real-time/Schwinger-Keldysh formalism, which we will review in the note.

Our goal in this note is to study an open EFT for interacting fermions via real-time holography \cite{Son:2002sd, Skenderis:2008dg, Skenderis:2008dh, Herzog:2002pc, Son:2009vu}. Real-time holography, in this context, refers to a method similar to the Gibbons-Hawking prescription \cite{Gibbons:1976ue} for computing real-time observables in gravitational systems. More precisely, there exists a conjectured geometry \cite{Glorioso:2018mmw} that dominates the real-time/Schwinger-Keldysh (SK) gravitational path integral. This geometry, known as \textit{grSK}, involves connecting two copies of the black hole exterior along their future horizons. Such advances in real-time holography have facilitated the development of open EFTs, albeit limited to free probe fermions \cite{Giecold:2009tt, Loganayagam:2020eue, Loganayagam:2020iol}. However, the inclusion of interactions is essential as probe fermions are not free in general.

From the bulk perspective, this study examines the scattering of fermions against a black hole background, considering the often overlooked influence of \textit{Hawking radiation} \cite{Hawking:1975vcx}. The grSK geometry accurately accounts for the thermal fluctuations of a black hole, i.e., Hawking radiation. Its validity across various scenarios has been demonstrated by numerous studies \cite{Chakrabarty:2019aeu, Jana:2020vyx, Loganayagam:2020eue, Loganayagam:2020iol, Colin-Ellerin:2020mva, Ghosh:2020lel, Bu:2020jfo, He:2021jna, He:2022jnc, He:2022deg, Loganayagam:2022zmq, Pantelidou:2022ftm, Loganayagam:2022teq, Loganayagam:2023pfb}. This framework successfully extends even to the cases with chemical potentials, such as charged \cite{Loganayagam:2020iol} and rotating black holes \cite{Chakrabarty:2020ohe}. Moreover, it inherently embodies the microscopic unitarity and the thermality of holographic systems. In \cite{Loganayagam:2024mnj}, it has been shown that the grSK results for boundary correlators align with Witten diagrammatics, suggesting an \emph{exterior field theory} for scalar theories without any derivative interactions. It is worth noting that grSK inherently provides an exterior field theory, a presumption made in earlier studies \cite{Liu:2009dm, Faulkner:2009wj, Iqbal:2011ae}. We aim to expand this investigation by including fermionic fields.

To simplify the analysis, we work with a background geometry having a planar horizon. More precisely, the AdS$_{d+1}$ Schwarzschild blackbrane grSK geometry. Furthermore, we assume that both the scalar and the spinor fields are massless, interacting via a Yukawa term. The dynamics of these fields in the grSK geometry can be obtained perturbatively (in the Yukawa coupling). This perturbative solution can then be used to compute the on-shell action, which is the leading contribution\footnote{Here we are referring to the leading contribution in the large $N$ expansion of the boundary theory.} to the boundary SK generating functional. 

Finally, it is important to highlight a key aspect of this note regarding our approach to presenting results. Specifically, we express the SK generating functional (or real-time correlators) in the column vector representation \cite{Henning:1993gh, Carrington:2006xj, Chaudhuri:2018ymp}. The column vector representation closely resembles the \textit{K\"{a}ll\a'{e}n-Lehmann} spectral representation extended to thermal scenarios. This representation not only reveals the intrinsic unitarity and thermality (Kubo-Martin-Schwinger conditions \cite{Kubo:1957mj, Martin:1959jp}) of the theory but also naturally includes all statistical factors (Bose-Einstein and Fermi-Dirac). While the scalar column vector representation has been explored previously, there has not been a generalization to spinors. In this note, we demonstrate that the results from grSK, including spinors (specifically Yukawa interactions), naturally align with a column vector representation. We conclude this introduction by providing a brief outline of the subsequent sections.

\subsection*{Outline}

A brief outline of this article is as follows. In Sec.~(\ref{sec:IFandgrSK}), we begin with a review of the influence functional and its connection to the Schwinger-Keldysh formalism. Here, we also briefly explain the gravitational counterpart of this formalism. In Sec.~(\ref{sec:YukawagrSK}), we consider Yukawa theory in the gravitational Schwinger-Keldysh geometry and solve for its dynamics in a perturbative expansion. In Sec.~(\ref{sec:YukawaandExteriorfromgrSK}), we explicitly calculate the three-point functions, and further present the Feynman rules that come from grSK. Using these Feynman rules, we list all the Witten diagrams that are relevant for the three-point and the four-point functions. Furthermore, we present the expressions for the generating functional explicitly in the column vector representation in Sec.~(\ref{sec:YukawaIF}). We conclude in Sec.~(\ref{sec:Discussion}) with a summary and a discussion of future directions. The appendices provide some technical details and explicit expressions omitted from the main text.

\section{Influence Functional and grSK geometry}\label{sec:IFandgrSK}

As discussed in the introduction, we aim to construct an open EFT for a \textit{system} of interacting bosonic and fermionic degrees of freedom coupled to a holographic \textit{thermal bath}. A way of obtaining such an EFT was first given by Feynman and Vernon \cite{Feynman:1963fq}, sometimes dubbed as the \textit{Influence functional formalism}. The basic idea is to double the degrees of freedom to describe the non-unitary evolution of the \textit{system}, which arises by integrating out the \textit{bath}.

The set-up we have is the following: consider a single system degree of freedom $\mathfrak{X}$ coupled with many bath degrees of freedom $X_i$. The total action of the combined unitary theory is
\begin{equation}
    S_{\text{tot}}[\mathfrak{X},X_i]=\ S_{\text{s}}[\mathfrak{X}]+S_{\text{b}}[X_i]+S_{\text{s-b}}[\mathfrak{X},X_i] \ .
\end{equation}
where $S_{\text{s}}$, $S_{\text{b}}$ and $S_{\text{s-b}}$ are parts of the action describing the system, the bath and the interaction between them respectively. Integrating out the bath degrees of freedom $X_i$, we obtain the non-unitary theory for the system. In the influence functional language, the path integral heuristically becomes
\begin{equation}
    \int \mathcal{D}\mathfrak{X} \, \mathcal{D}X_i \,  \exp \bigg( i S_{\text{tot}}[\mathfrak{X},X_i] \bigg) \mapsto \int \mathcal{D}\mathfrak{X}_{\sR} \, \mathcal{D}\mathfrak{X}_{\sL} \exp \bigg( i S_{\text{s}}[\mathfrak{X}_{\sR}]-i S_{\text{s}}[\mathfrak{X}_{\sL}]+i S_{\text{IF}}[\mathfrak{X}_{\sR},\mathfrak{X}_{\sL}] \bigg) \ ,
\end{equation}
where $\mathfrak{X}_{\sR}$ and $\mathfrak{X}_{\sL}$ are two copies of system degrees of freedom and  $S_{\text{IF}}[\mathfrak{X}_{\sR},\mathfrak{X}_{\sL}]$ is the \textit{influence phase}, containing the influence of the bath on the system.

The influence phase generates the real-time correlators of the operator $\mathcal{O}(X_i)$ coupled to the system degree of freedom $\mathfrak{X}$.\footnote{Here we assume that the interaction action $S_{\rm s-b}$ is bilinear in the operators $\mathcal{O}(X_i)$ and $\mathfrak{X}$} More precisely, $S_{\text{IF}}[\mathfrak{X}_{\sR},\mathfrak{X}_{\sL}]$ is the real-time or the \textit{Schwinger-Keldysh} generating functional of the bath. The Schwinger-Keldysh (SK) formalism is the appropriate real-time framework for addressing non-equilibrium questions in QFT \cite{kamenev_2011,Bellac:2011kqa,Rammer:2007zz}. This is especially true if we want to systematically account for fluctuations.

\begin{figure}[H]
\centering
\begin{tikzpicture}[scale=1.6,decoration={markings,
mark=at position 8cm with {\arrow[line width=1pt]{>}},
mark=at position 15cm with {\arrow[line width=1pt]{>}};
}
]
\draw[white!20!gray, line width = .7] (1,1.2) -- (1,-0.1) ;
\draw[white!20!gray, line width = .7] (0.4,1) --(9,1);
\filldraw [black] (1,1) circle (1pt); 
\filldraw [black] (1,0) circle (1pt);  
\path[draw,teal!85!green,line width=1.5pt,postaction=decorate](1,1) -- (8.5,1) -- (8.5,0.8)  -- (1,0.8)  -- (1,0);
\draw (0.8,1.1)  node[black] {\scriptsize $A$};
\draw (0.8,0)  node[black] {\scriptsize $B$};
\draw (5.8,1.2)  node[black] {$\mathcal{C}$};
\draw (3.8,1.2)  node[black] {\scriptsize{(R)}};
\draw (3.8,0.5)  node[black] {\scriptsize(L)};
\draw (1.2,1.2)  node[black] {\scriptsize$t_i$};
\draw (8.7,1.2)  node[black] {\scriptsize$t_f$};
\draw (1.4,0)  node[black] {\scriptsize$t_i -i \beta$};
\draw (8.8,0.5)  node[black] {\scriptsize$t_f-i \epsilon$};
\end{tikzpicture}
\caption{The SK contour at finite temperature $T=\frac{1}{\beta}$. The direction of the arrow represents the direction of contour time $t_{\mathcal{C}}$, involving both forward (R) and backward (L) time-evolving parts. Note that points $A$ and $B$ are identified in the contour.}
\label{Fig:SKcontour}
\end{figure}
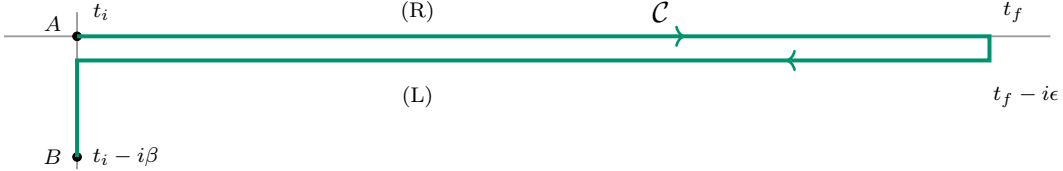

In SK formalism, we have a path-integral that evolves mixed states: this is done through a closed time contour with a forward and backward time evolution. Such a contour allows us to directly compute real-time correlators. In Fig.~(\ref{Fig:SKcontour}), we depict the SK contour for a thermal state in the complex time plane.
To be more concrete, consider a holographic conformal field theory in $d$  spacetime dimensions (CFT$_{d}$), described by the action $S_{\text{CFT}}$. Let $\mathcal{O}$ be a bosonic operator in this CFT, and $J$ be its source, both living on the SK contour $\mathcal{C}$. The SK generating functional $\mathcal{Z}_{\text{SK}}$ of the CFT is then given by the path-integral 
\begin{equation}
    \mathcal{Z}_{\text{SK}} \left[J\right]=  \int \mathcal{D} \mathcal{O} \   \exp \bigg( i S_{\text{CFT}} +i \oint \diff t_{_{\mathcal{C}}} \diff^{d-1} \textbf{x} \ J(x) \mathcal{O}(x)\bigg)  \ .
\end{equation}
Assuming there are no sources on the imaginary part of the contour,\footnote{We have not put any sources on the imaginary part of the contour as it would change the initial thermal state.} it is convenient to write the fields on the right and the left parts separately. To do so, we introduce the following notation:
\begin{equation}
    \begin{aligned}
    & \mathcal{O}(t) \equiv \ \mathcal{O}_{\sR}(t)\ ,  &&\quad  J(t) \equiv\ J_{\sR}
    (t) &&& \forall \quad t \in \mathbb{R} \ , \\
    & \mathcal{O}(t-i\epsilon)\equiv\ \mathcal{O}_{\sL}(t)\  ,
    && \quad J(t-i\epsilon) \equiv\ J_{\sL}(t)  &&&   \forall \quad t \in \mathbb{R}  \ ,
    \end{aligned}
\end{equation}
where the subscript denotes the part of the contour on which the field lives. We can then write
\begin{equation}
    \mathcal{Z}_{\text{SK}}[J_{\sR},J_{\sL}]= \Bigg \langle \exp \Bigg\{i \int \diff^d x \  \Big[ J_{\sR}(x) \mathcal{O}_{\sR}(x)-J_{\sL}(x) \mathcal{O}_{\sL}(x) \Big]  \ \Bigg\} \Bigg \rangle_{\rm CFT}    \ . 
\end{equation}
By taking the functional derivatives of $\mathcal{Z}_{\text{SK}}$, we obtain two-point Schwinger-Keldysh correlators $K_{_{AB}}$ as follows,
\begin{equation}
    i K_{_{AB}}(t_1,t_2) =\ \frac{1}{i^2} \frac{\delta^2 \, \log \mathcal{Z}_{\text{SK}}}{\delta J_{_{A}}(t_1)\delta J_{_{B}}(t_2)}\Bigg|_{J=0} \ .
\end{equation}

In the operator formalism, these SK correlators correspond to the contour-ordered correlation functions. In contour ordering, time increases along the arrow on the upper part of the contour and decreases along the arrow on the lower part. Moreover, any point on the lower part is considered to have greater \textit{contour time} than any point on the upper part. Then, it is not hard to check that the following holds:
\begin{equation}
    \begin{split}
        &iK_{_{RR}}(t_1,t_2) =\  \braket{\mathbb{T}\{\mathcal{O}(t_1)\mathcal{O}(t_2)\}}\ ,\qquad
        iK_{_{RL}}(t_1,t_2)  =\  \braket{\mathcal{O}(t_2)\mathcal{O}(t_1)} \ ,   \\
        & iK_{_{LL}}(t_1,t_2) =\ \braket{\overline{\mathbb{T}}\{\mathcal{O}(t_1)\mathcal{O}(t_2)\}}\ , \qquad
        iK_{_{LR}}(t_1,t_2)  =\  \braket{\mathcal{O}(t_1)\mathcal{O}(t_2)}\ .
    \end{split}
\end{equation}
where $\mathbb{T}$ and $\overline{\mathbb{T}}$ represent the time ordering and the anti-time ordering in real-time, respectively. 

The SK formalism is robust as it applies from free to strongly interacting systems. It makes manifest fundamental principles like unitarity, causality and thermality. The SK 2-point correlators naturally encode the retarded, advanced and fluctuation responses of the field theory. Here, we have only discussed the bosonic SK formulation but the whole formalism can be easily generalized to fermions. Without going into the details of all of this,\footnote{For a detailed discussion on SK formalism, we recommend the readers to look to textbooks on non-equilibrium QFT \cite{kamenev_2011,Bellac:2011kqa,Rammer:2007zz}.} we now turn our attention to the gravitational counterpart of the SK formalism.   

In the gravitational dual, the thermal states of the Conformal Field Theory (CFT) correspond to AdS black holes. The duplication of CFT fields directly corresponds to the doubling of the AdS black hole geometry \cite{Skenderis:2008dg,Skenderis:2008dh}. The \textit{gravitational Schwinger-Keldysh (grSK) geometry} should be constructed from these two copies of the AdS black hole geometry such that it gives the SK contour in the boundary limit \cite{Glorioso:2018mmw}. Hence, the SK generating functional $\mathcal{Z}_{\text{SK}}$ can be computed by evaluating the on-shell gravitational action \cite{Chakrabarty:2019aeu,Jana:2020vyx}.

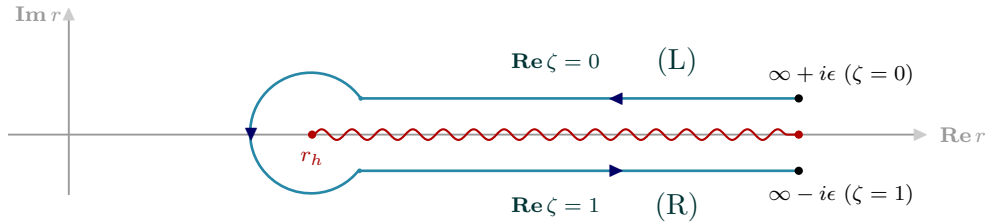
\begin{figure}[H]
\centering
	\begin{tikzpicture}[scale=0.8]
	\draw[white!20!gray, line width = .7] (-4,2) -- (-4,-1) ;
	\draw[white!20!gray, line width = .7] (-5,0) --(10,0);
	\draw[cyan!20!teal, line width = 1] (0.8,0.6) -- (8,0.6);
	\draw[cyan!20!teal, line width = 1] (0.8,0.6) arc (35: 322 : 1);
	\draw[cyan!20!teal, line width = 1] (0.79,-0.6) --(8,-0.6);
	\draw[decoration={snake, segment length = 4 mm, amplitude=2}, decorate, red!70!black, thick] (0,0) -- (8,0);
	
	\node[circle,scale= 0.3,red!70!black,fill] at (0,0) {} ;
	\node[circle,scale= 0.3,red!70!black,fill] at (8,0) {} ;
	\node[circle, fill, cyan!20!teal, scale= 0.17] at (0.8,0.6) {} ;
	\node[circle, fill, cyan!20!teal, scale= 0.17] at (0.78,-0.6) {};
	\node[circle, fill, black, scale= 0.3] at (8,0.6) {} ;
	\node[circle, fill, black, scale= 0.3] at (8,-0.6) {} ;
	
	\node[white!60!gray] at (-4,2) {\scriptsize$\blacktriangle$};
	\node[white!60!gray] at (10,0) {\scriptsize $\blacktriangleright$};
	\node[blue!40!black] at (5,0.6) {\scriptsize $\blacktriangleleft$};
	\node[blue!40!black] at (5,-0.6) {\scriptsize $\blacktriangleright$};
	\node[blue!40!black] at (-1,0) {\scriptsize $\blacktriangledown$};
	
	\node[teal!40!black] at (6,1.2) {(L)};
	\node[teal!40!black] at (6,-1.2) {(R)};	
	
	\node[teal!40!black] at (4,1.2) {\scriptsize $\mathbf{Re} \,  \zeta = 0$};
	\node[teal!40!black] at (4,-1.2) {\scriptsize $\mathbf{Re} \, \zeta  = 1$};
	\node[red!65!black] at (0,-0.4) {\scriptsize $r_h$};
	
	\node at (8.7,1) {\scriptsize $\infty+i\epsilon \ (\z=0)$};
	\node at (8.7,-1) {\scriptsize $\infty-i\epsilon \ (\z=1)$};
	\node[white!20!gray] at (-4.5, 2) {\scriptsize $\mathbf{Im} \, r$};
	\node[white!20!gray] at (10.7,0) {\scriptsize $\mathbf{Re} \, r$};
	\end{tikzpicture}
	\caption{ \textbf{ grSK geometry : }The radial contour drawn on the complex $r$ plane at fixed $v$, parametrized by coordinate $\z$.}
	\label{grSKcontour}
\end{figure}

The grSK geometry is obtained by glueing two copies of the blackbrane exterior across their future horizons. Consider a blackbrane in asymptotically AdS$_{d+1}$ background in \textit{ingoing Eddington-Finkelstein} (EF) coordinates given by
\begin{equation}\label{eq:ingoingEF}
\diff s^2 =\  - r^2 f(r) \, \diff v^2 + 2\,\diff v \,  \diff r + r^2  \diff \textbf{x}^2 \ , \qquad \qquad f(r)=1-\left( \frac{r_{h}}{r}\right)^d  .
\end{equation}
where $r_{h}$ is the horizon radius. We first complexify the radial coordinate resulting in a $(d+2)$-real-dimensional manifold. The grSK geometry is then defined as a codimension-1 slice of this manifold: the radial coordinate in grSK varies along a contour in the  complex radial plane as shown in Fig.~(\ref{grSKcontour}).
We will find it convenient to parametrize this radial contour  by a complexified tortoise co-ordinate $\z$ defined via
\begin{equation}
\frac{\diff \z}{\diff r } = \frac{2}{i \beta \, r^{2} f(r)} \ ,   
\end{equation}
where $ \beta \equiv \frac{4 \pi}{d r_{h}}$ is the inverse Hawking temperature of the black brane. The coordinate $\z$ has a branch cut along the exterior and is
normalised to have a unit monodromy around the horizon branch point $r=r_{h}$.

\begin{figure}[H]
    \centering
	\tikzset{every picture/.style={line width=0.75pt}} 
	
	\begin{tikzpicture}[scale=1.4,x=0.75pt,y=0.75pt,yscale=-1,xscale=1]
	
	\draw  [dash pattern={on 4.5pt off 4.5pt}]  (79.33,39.67) -- (78.67,119.67) ;
	\draw    (158.67,41) -- (158,121) ;

	\draw  [dash pattern={on 4.5pt off 4.5pt}]  (78.67,119.67) -- (158,121) ;
	
	\draw    (79.33,39.67) .. controls (81.02,38.03) and (82.69,38.06) .. (84.33,39.75) .. controls (85.97,41.44) and (87.64,41.47) .. (89.33,39.83) .. controls (91.02,38.2) and (92.69,38.23) .. (94.33,39.92) .. controls (95.97,41.61) and (97.64,41.64) .. (99.33,40) .. controls (101.02,38.37) and (102.69,38.4) .. (104.33,40.09) .. controls (105.97,41.78) and (107.64,41.81) .. (109.33,40.17) .. controls (111.02,38.53) and (112.69,38.56) .. (114.33,40.25) .. controls (115.97,41.94) and (117.64,41.97) .. (119.33,40.34) .. controls (121.02,38.7) and (122.69,38.73) .. (124.33,40.42) .. controls (125.97,42.11) and (127.64,42.14) .. (129.33,40.51) .. controls (131.02,38.87) and (132.69,38.9) .. (134.33,40.59) .. controls (135.96,42.28) and (137.63,42.31) .. (139.32,40.67) .. controls (141.01,39.04) and (142.68,39.07) .. (144.32,40.76) .. controls (145.96,42.45) and (147.63,42.48) .. (149.32,40.84) .. controls (151.01,39.21) and (152.68,39.24) .. (154.32,40.93) -- (158.67,41) -- (158.67,41) ;

	\draw    (79.33,39.67) -- (158,121) ;
	\draw    (158.67,41) -- (118.67,80.33) (152.94,49.44) -- (150.13,46.59)(145.81,56.45) -- (143,53.6)(138.68,63.46) -- (135.87,60.61)(131.55,70.47) -- (128.74,67.62)(124.42,77.48) -- (121.61,74.63) ;
	\draw  [fill={rgb, 255:red, 74; green, 74; blue, 74 }  ,fill opacity=1 ] (187.33,131.13) -- (191.5,131.13) -- (191.5,112.33) -- (199.83,112.33) -- (199.83,131.13) -- (204,131.13) -- (195.67,143.67) -- cycle ;
	\draw    (150,171.67) -- (156.67,288.33) ;
	\draw    (150,171.67) -- (139.33,293.67) ;
	\draw    (150,171.67) -- (80,233.67) (143.84,179.79) -- (141.19,176.8)(136.35,186.42) -- (133.7,183.43)(128.87,193.05) -- (126.22,190.06)(121.38,199.69) -- (118.73,196.69)(113.9,206.32) -- (111.24,203.32)(106.41,212.95) -- (103.76,209.95)(98.92,219.58) -- (96.27,216.58)(91.44,226.21) -- (88.79,223.21)(83.95,232.84) -- (81.3,229.84) ;
	\draw    (80,233.67) -- (139.33,293.67) ;

	\draw  [dash pattern={on 0.84pt off 2.51pt}]  (80,233.67) -- (156.67,288.33) ;
	\draw  [color={rgb, 255:red, 74; green, 144; blue, 226 }, dash pattern={on 1.84pt off 2.51pt}]  (108,212.67) -- (146.67,238.33) ;

	\draw [color={rgb, 255:red, 144; green, 19; blue, 254 }  ,draw opacity=1 ]   (132,67) -- (158.5,97) ;
	\draw [color={rgb, 255:red, 144; green, 19; blue, 254 }  ,draw opacity=1 ]   (144,56) -- (158.33,71) ;

	\draw  [dash pattern={on 4.5pt off 4.5pt}]  (222.67,41) -- (222,121) ;
	\draw    (302,42.33) -- (301.33,122.33) ;
	\draw  [dash pattern={on 4.5pt off 4.5pt}]  (222,121) -- (301.33,122.33) ;
	
\draw    (222.67,41) .. controls (224.36,39.36) and (226.03,39.39) .. (227.67,41.08) .. controls (229.31,42.77) and (230.98,42.8) .. (232.67,41.17) .. controls (234.36,39.53) and (236.03,39.56) .. (237.66,41.25) .. controls (239.3,42.94) and (240.97,42.97) .. (242.66,41.34) .. controls (244.35,39.7) and (246.02,39.73) .. (247.66,41.42) .. controls (249.3,43.11) and (250.97,43.14) .. (252.66,41.5) .. controls (254.35,39.87) and (256.02,39.9) .. (257.66,41.59) .. controls (259.3,43.28) and (260.97,43.31) .. (262.66,41.67) .. controls (264.35,40.04) and (266.02,40.07) .. (267.66,41.76) .. controls (269.3,43.45) and (270.97,43.48) .. (272.66,41.84) .. controls (274.35,40.2) and (276.02,40.23) .. (277.66,41.92) .. controls (279.3,43.61) and (280.97,43.64) .. (282.66,42.01) .. controls (284.35,40.37) and (286.02,40.4) .. (287.66,42.09) .. controls (289.3,43.78) and (290.97,43.81) .. (292.66,42.18) .. controls (294.35,40.54) and (296.02,40.57) .. (297.66,42.26) -- (302,42.33) -- (302,42.33) ;

	\draw    (222.67,41) -- (301.33,122.33) ;

	\draw    (302,42.33) -- (262,81.67) (296.27,50.77) -- (293.47,47.92)(289.14,57.78) -- (286.34,54.93)(282.01,64.79) -- (279.21,61.94)(274.88,71.8) -- (272.08,68.95)(267.75,78.82) -- (264.95,75.96) ;

	\draw [color={rgb, 255:red, 74; green, 144; blue, 226 }  ,draw opacity=1 ]   (274,69) -- (302,98) ;
	\draw [color={rgb, 255:red, 74; green, 144; blue, 226 }  ,draw opacity=1 ]   (287,57) -- (301.67,72.33) ;

	\draw [color={rgb, 255:red, 144; green, 19; blue, 254 }  ,draw opacity=1 ]   (104.67,211.67) -- (143,250.33) ;

	\draw [color={rgb, 255:red, 74; green, 144; blue, 226 }  ,draw opacity=1 ]   (144,236.5) -- (154,244) ;

    \draw [color={rgb, 255:red, 0; green, 0; blue, 0}  ,draw opacity=1 ]   (142,277.8) -- (156.9,288.5) ;

\draw [shift={(152.78,218.73)}, rotate = 266.58] [fill={rgb, 255:red, 0; green, 0; blue, 0 }  ][line width=0.08]  [draw opacity=0] (3.57,-1.72) -- (0,0) -- (3.57,1.72) -- cycle    ;

\draw [shift={(145.8,222.11)}, rotate = 95.23] [fill={rgb, 255:red, 0; green, 0; blue, 0 }  ][line width=0.08]  [draw opacity=0] (3.57,-1.72) -- (0,0) -- (3.57,1.72) -- cycle    ;

\draw [shift={(125.91,233.7)}, rotate = 227.87] [fill={rgb, 255:red, 144; green, 19; blue, 254 }  ,fill opacity=1 ][line width=0.08]  [draw opacity=0] (3.57,-1.72) -- (0,0) -- (3.57,1.72) -- cycle    ;

\draw [shift={(130.2,227.3)}, rotate = 37.92] [fill={rgb, 255:red, 74; green, 144; blue, 226 }  ,fill opacity=1 ][line width=0.08]  [draw opacity=0] (3.57,-1.72) -- (0,0) -- (3.57,1.72) -- cycle    ;
	
	\draw (22.67,14.67) node [anchor=north west][inner sep=0.75pt]   [align=left] {{ \textbf{{\footnotesize 1. Take two AdS Schwarzschild blackbranes in ingoing EF coordinates.}}}};
	\draw (22.67,146.33) node [anchor=north west][inner sep=0.75pt]   [align=left] {{ \textbf{{\footnotesize 2. Stitch them across their future horizons, i.e. $\mathcal{H}^+$.}}} };
	\draw (98.33,188.67) node [anchor=north west][inner sep=0.75pt]   [align=left] {{\scriptsize $\mathcal{H}^+$}};

	\draw (117.35,50) node [anchor=north west][inner sep=0.75pt]   [align=left] {{\fontsize{0.57em}{0.68em}\selectfont $\mathcal{H}^+$}};
	\draw (90.67,261.33) node [anchor=north west][inner sep=0.75pt]   [align=left] {{\scriptsize $\mathcal{H}^-$}};
	\draw (117.67,93.67) node [anchor=north west][inner sep=0.75pt]   [align=left] {{\fontsize{0.57em}{0.68em}\selectfont $\mathcal{H}^-$}};
	\draw (260.67,97.67) node [anchor=north west][inner sep=0.75pt]   [align=left] {{\fontsize{0.57em}{0.68em}\selectfont $\mathcal{H}^-$}};
	\draw (260.33,53.33) node [anchor=north west][inner sep=0.75pt]   [align=left] {{\fontsize{0.57em}{0.68em}\selectfont $\mathcal{H}^+$}};
		
	\end{tikzpicture}
	\caption{ Schematic representation of the construction of the grSK geometry.}
	\label{fig:grSK1}
\end{figure}
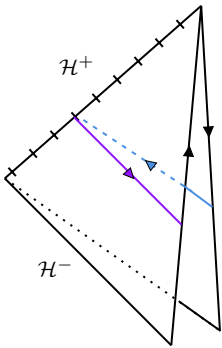

To gain a clearer insight into the grSK geometry, it is helpful to use Penrose diagrams. These diagrams provide a clear visualization of how the grSK geometry is constructed from the standard black brane geometry. The grSK geometry is constructed in two steps, illustrated in Fig.~(\ref{fig:grSK1}). 
Here, we first consider the two exteriors of the AdS blackbrane geometry in ingoing EF coordinates. Next, we connect these geometries across their future horizons, resulting in the formation of the grSK geometry.

Regarding the diagram in Fig.~(\ref{fig:grSK1}), it is important to note a couple of aspects. Firstly, the diagram does not explicitly depict what happens at the future horizons during the stitching process. Additionally, it lacks a connection to the \textit{cigar} geometry, which represents the initial thermal state in the CFT. Therefore, it is crucial to view this Penrose diagram of the grSK geometry as purely representational.\footnote{For a more in-depth understanding of the Penrose diagram of the grSK geometry, readers are encouraged to look at \cite{Jana:2020vyx}.} Concluding our discussion of the grSK geometry, it is worth mentioning that this framework has been extended to include charged and rotating black holes, as explored in \cite{Loganayagam:2020iol,Chakrabarty:2020ohe}.

Having clarified the background geometry, the next question is: how do different kinds of fields propagate and interact in this background? This question has already been explored for a variety of holographic systems \cite{Chakrabarty:2019aeu, Jana:2020vyx, Loganayagam:2020eue,Loganayagam:2020iol, Colin-Ellerin:2020mva, Ghosh:2020lel, Bu:2020jfo, He:2021jna, He:2022jnc, He:2022deg, Loganayagam:2022zmq, Pantelidou:2022ftm,Loganayagam:2022teq,Loganayagam:2023pfb,Chakrabarty:2020ohe,Loganayagam:2024mnj}. However, for the first time, we aim to investigate systems involving interacting fermions. As an illustrative example in this context, we will consider both bosons and fermions interacting through the Yukawa interaction.

\section{Yukawa theory in grSK}\label{sec:YukawagrSK}

We will now consider one of the simplest theories involving Dirac spinors, i.e.,  Yukawa theory. For simplicity, here we will be concerned with a massless Dirac spinor interacting with a massless scalar through the Yukawa interaction. Furthermore, both of these will be minimally coupled to the grSK spacetime, which itself will be non-dynamical.

Yukawa interactions in the bulk are described by the action
\begin{equation}
S = S_{\rm free} + S_{\rm int}  \ ,
\end{equation}
where $S_{\rm free}$ is the sum of the free massless Dirac action and the free massless scalar action,
and $S_{\rm int}$ is the Yukawa interaction term defined as
\begin{equation}
S_{\rm int} = - i\oint \diff \zeta \int \diff^d x  \sqrt{-g} \ \lambda \phi \bar{\Psi} \Psi \ .
\end{equation}
Here $\lambda$ is the Yukawa coupling constant that controls the strength of the coupling between the Dirac spinor $\Psi$ and the scalar $\phi$. The symbol $g$ denotes the determinant of the grSK metric. The explicit expression of the free part of the above action will be described in detail below.

Before proceeding to understand this action and its consequences, we will first focus on the free part. We will describe scalar and spinor propagation in the grSK geometry without any interaction terms between them. This will also allow us to set up necessary notation for the later sections. Later in this section, as well as in the next section, we will use this notation to describe the Yukawa interaction in detail.

\subsection{Free scalar theory}
We start with scalar propagation in the grSK geometry, i.e., we consider the scalar part of the free action given above. We will see how to solve the Klein-Gordon equation that comes from varying this action. The important lesson here is that once we have one solution to the Klein-Gordon equation, we then can use the time-reversal involution \cite{Loganayagam:2020eue} of the grSK geometry to generate another independent solution. We will see this in more detail soon.  

The free scalar field is described by the action
\begin{equation}
S_{\phi}=\ - \oint \diff \zeta \int  \mathrm{d}^{d} x\sqrt{-g}\ \frac{1}{2}g^{M N}\partial_{M}\phi \partial_{N}\phi  \ ,
\end{equation}
where $g^{MN}$ is the metric describing the grSK spacetime, and $g$ is the determinant of this metric. Varying the above action, we obtain the equation of motion
\begin{equation}
    \frac{1}{\sqrt{-g}} \partial_{M} \left(\sqrt{-g} g^{M N} \partial_{N}\right) \phi = 0 \ .
\end{equation}
The above partial differential equation can be reduced to a second-order ordinary differential equation in the radial coordinate by passing to the Fourier domain in the boundary coordinates. In other words, if we decompose the field $\phi$ as
\begin{equation}
    \phi(\zeta, x) \equiv \int_{k} e^{i k x} \phi(\zeta, k)\ , \qquad \int_{k}  \equiv \int \frac{\diff^d k}{(2\pi)^d} \ ,
    \label{eq:DefFourTransScalar}
\end{equation}
then we obtain the equation of motion
\begin{equation}
    \mathbb{D}_{+} \left(r^{d-1} \mathbb{D}_+ \phi\right) + \frac{\beta^2}{4}r^{d-1} \left(f |\bk|^2  -(k^0)^2\right)\phi = 0\ ,
\end{equation}
where $\mathbb{D}_{+}$ is the radial derivative operator defined as
\begin{equation}
    \mathbb{D}_{+} \equiv \frac{\diff }{\diff \zeta} + \frac{\beta k^0}{2}\ .
\end{equation}
Note that we use the same symbol for the field in the real domain as well as the Fourier domain. This should not cause any confusion since the arguments of the field will differentiate between them.

The above second-order ordinary differential equation has two independent solutions. The solution $G^{\rm in}(\zeta,k)$ that satisfies the boundary conditions
\begin{equation}
    G^{\rm in}(\zeta,k)\Big|_{\zeta = 0 } = 1 = G^{\rm in}(\zeta,k)\Big|_{\zeta = 1}
\end{equation}
is called the \emph{ingoing boundary-to-bulk} Green function. Given this ingoing solution, we can generate another independent solution, the \emph{outgoing boundary-to-bulk} Green function, by using the time-reversal involution of the grSK geometry
\begin{equation}
    G^{\rm out}(\zeta,k) = e^{-\beta k^0 \zeta} G^{\rm in}(\zeta, -k) \ .
\end{equation}

With these two independent solutions, the most general solution of the Klein-Gordon equation satisfying the boundary conditions
\begin{equation}
    \phi(\zeta,k) \Big|_{\zeta=0} = J_{\sL}(k) \ , \quad \phi(\zeta,k) \Big|_{\zeta=1} = J_{\sR}(k) \ ,
\end{equation}
takes the form
\begin{equation}
    \phi(\zeta,k) = -G^{\rm in}(\zeta, k) J_{\Fb}(k) + e^{\beta k^0} G^{\rm out} J_{\Pb}(k) \ ,
\end{equation}
where $J_{\Fb}$ and $J_{\Pb}$ are the future and the past sources defined as 
\begin{equation}
    \begin{split}
        J_{\Fb}(k) &\equiv - (1+n^{\be}_{k^0}) J_{\sR}(k) + n^{\be}_{k^0} J_{\sL}(k)\ ,\\
        J_{\Pb}(k) &\equiv -n^{\be}_{k^0} \left[J_{\sR}(k) - J_{\sL}(k)\right] \ ,
    \end{split}
\end{equation}
respectively. This form of the solution with the ingoing and the outgoing propagators was first explored in \cite{Son:2009vu} and later detailed for the grSK geometry in \cite{Chakrabarty:2019aeu, Jana:2020vyx}. 
The symbol $n^{\be}_{k^0}$ denotes the Bose-Einstein statistical factor
\begin{equation}
    n^{\be}_{k^0} \equiv \frac{1}{e^{\beta k^0} -1}\ .
\end{equation}

We will also later find it convenient to write the above solution in the right-left basis of the sources as
\begin{equation}\label{Phi0main}
\begin{split}
\phi(\zeta, k) &\equiv \   g_{\sR}(\zeta,k)\,J_{\sR}(k)-g_{\sL}(\zeta,k)\,J_{\sL}(k) \ ,
\end{split}
\end{equation}
where $g_{\sR,\sL}$ denote the boundary-to-bulk Green functions from the right and the left AdS boundaries, which in terms of the ingoing and the outgoing Green functions take the form 
\begin{equation}\label{gRgLmain}
\begin{split}
&g_{\sR}(\zeta,k)  \equiv  \left(1+n^{\be}_{k^0}\right)\left[G^{\text{in}}(\zeta,k)-G^{\text{out}}(\zeta,k)  \right]\ ,  \\
&g_{\sL}(\zeta,k) \equiv n^{\be}_{k^0}\left[G^{\text{in}}(\zeta,k)-e^{\beta k^0 }G^{\text{out}}(\zeta,k) \right]  \ .
\end{split}
\end{equation}

\subsection{Free Dirac theory}
Next, we come to free Dirac spinor propagation in the grSK geometry. The free Dirac action takes the form
\begin{equation}\label{eq:Diracaction}
    S_{\Psi} = \ i \oint \diff \zeta \int \diff^d x \sqrt{-g}\, \bar{\Psi}\G^{\A}D_{\A} \Psi +i\int \diff^d x \, \bar{\Psi} \Pmin \Psi\Bigg|_{\z=0}^{\z=1} \ .
\end{equation}
Notice that the first term is just the familiar massless Dirac action in curved spacetime, where the symbol $D_{\A}$ represents the appropriate covariant derivative, and $\mathbb{P}_{\pm}$ are the appropriate projection operators. Explicitly,
\begin{equation}
    D_{\A} \equiv \partial_{\A}+\frac{1}{4}\w_{ab \A}\G^{a}\G^{b} \ , \qquad \mathbb{P}_{\pm}\equiv \frac{1}{2}\left(\mathbb{I} \pm \Gamma^{(\zeta)}\right) \ .
    \label{eq:CovDerDiracSpinors} 
\end{equation}
Here $\omega_{abA}$ are the spin connection coefficients and $\Gamma^a$ are the bulk gamma matrices.\footnote{Our notation for the indices for the gamma matrices is as follows: Lowercase Latin indices are for vielbein frame gamma matrices whereas uppercase Latin indices are for spacetime gamma matrices. More precisely, $\{\Gamma^{a}, \Gamma^{b}\} =  2 \eta^{ab}$ and $\{\Gamma^{\A}, \Gamma^{\B}\} =  2 g^{\A \B}$. When writing explicit components, we will denote vielbein frame gamma matrices with the superscript enclosed in parentheses. For example, $\Gamma^{(\zeta)}$. For more details about the conventions we use, the reader is encouraged to see \cite{Loganayagam:2020eue}. } Explicit expressions for these can be found in \cite{Loganayagam:2020eue}. The second term in the action above is a boundary term that ensures the existence of a well-defined variational principle. This is because spinors (as opposed to scalars) satisfy double semi-Dirichlet boundary conditions \cite{Henningson:1998cd, Henneaux:1998ch}. 

Varying the above action with respect to $\Psibar$ gives us the Dirac equation
\begin{equation}
    \Gamma^{\A} D_{\A} \Psi = 0 \ .
\end{equation}
Passing to the Fourier domain in the boundary coordinates, as we did for the scalar field in Eq.~\eqref{eq:DefFourTransScalar}, the Dirac equation becomes a first-order ordinary differential equation in the radial coordinate:
\begin{equation}
    \begin{split}
        \left\{ \frac{\bbT(\zeta)}{\sqrt{f}}\partial _\zeta + \mathbb{\Gamma} \left( \beta k^0 + \frac{1}{2} \partial_\zeta \ln f \right) - \frac{\beta}{2} \Gamma^{(i)} k_i  \right\}\left(r^{d/2} \Psi (\zeta,k) \right)  = 0 \ .
    \end{split}
    \label{eq:DiracFreeExplicitFourier}
\end{equation}
Here $\mathbb{\Gamma}$ and $\mathbb{T}$ are linear combinations of bulk gamma matrices defined as
\begin{equation}
    \mathbb{\Gamma} \equiv \frac{1}{2} \left(\Gamma^{(\zeta)}+\Gamma^{(v)}\right)  \ , \quad \bbT(\zeta) \equiv \sqrt{f} \mathbb{\Gamma} + \frac{1}{\sqrt{f}} \mathbb{\Gamma} \ .
\end{equation}

In the case of Dirac spinors as well, we would like to follow the same line of logic as we did for scalars to find the solutions. But here there are new important features we have to keep in mind. Primarily, the equation of motion is now a first-order equation. Secondly, the field we are solving for is a spinor, i.e., it has a spinorial index. 

Due to the latter feature, the boundary conditions of the bulk field are slightly more complicated. This is because in general dimensions, the boundary spinor and the bulk spinor do not have the same number of components. If $\psi_{\sR}$ and $\psi_{\sL}$ are spinors on the right and the left boundary respectively, the boundary conditions on the bulk spinor $\Psi$ take the form
\begin{equation}\label{eq:BCDiracSol}
    \lim_{\zeta \to 0}r^{\frac{d}{2}} \Pplus S_{_{0}}\Psi(\zeta,k) = \Pplus S_{_{0}}\psi_{\sL}(k) \ , \qquad \qquad \lim_{\zeta \to 1}r^{\frac{d}{2}} \Pplus S_{_{0}}\Psi(\zeta,k) = \Pplus S_{_{0}}\psi_{\sR}(k) \ .
\end{equation}
Here $S_0$ is a constant rectangular matrix defined in \cite{Loganayagam:2020eue}. Notice that since the Dirac equation is first-order in derivatives, the double Dirichlet boundary conditions can be imposed only after projecting with an appropriate projection matrix, which in the present case is $\projp$.

We can now proceed as in the scalar case by first introducing the ingoing boundary-to-bulk Green function $S^{\rm in}$, which is a solution of the Dirac equation and satisfies the boundary conditions
\begin{equation}\label{eq:BCSin}
    \lim_{\zeta \to 0}r^{\frac{d}{2}} \Pplus \Sin(\zeta,k)=  \Pplus S_{_{0}} =\lim_{\zeta \to 1}r^{\frac{d}{2}} \Pplus \Sin(\zeta,k)   \ .
\end{equation}
Given the ingoing Green function, once again the time-reversal involution of the grSK geometry produces outgoing Green function $S^{\rm out}$ given by
\begin{equation}
    S^{\rm out} (\zeta,k) \equiv e^{-\beta k^0 \zeta} \mathbb{T}(\zeta) S^{\rm in}(\zeta,-k) \ .
\end{equation}
The most general solution of the Dirac equation satisfying the boundary conditions in Eq.\eqref{eq:BCDiracSol} can then be written as
\begin{equation}
    \Psi(\zeta,k) = \ -\Sin(\zeta,k)\psiF(k)-e^{\beta k^0} \Sout(\zeta,k) \psiP(k) \ .
    \label{eq:FermionSonTeaney}
\end{equation}
Here $\psiF(k)$ and $\psiP(k)$ are the boundary sources in the past-future basis defined as
\begin{equation}
\psiF(k) \equiv \ n^{\fd}_{k^0}\left[\psi_{\sR}(k) -\psi_{\sL}(k)\right] -\psi_{\sR}(k)\  , \qquad \psiP(k) \equiv \ n^{\fd}_{k^0}\left[\psi_{\sR}(k) -\psi_{\sL}(k)\right] \ ,
\end{equation}
where $n^{\fd}_{k^0}$ is the Fermi-Dirac statistical factor
\begin{equation}
    n^{\fd}_{k^0} \equiv \frac{1}{e^{\beta k^0} +1}\ .
\end{equation}

We will also find it useful to recast the above solution in the right-left basis of the boundary sources:
\begin{equation}
    \Psi(\zeta, k) \equiv s_{\sR}(\zeta,k)\ \psi_{\sR}(k)-s_{\sL}(\zeta,k)\ \psi_{\sL}(k) \ 
\end{equation}
where $s_{\sR}$ and $s_{\sL}$ denote the boundary-to-bulk Green functions from the right and the left AdS boundaries, respectively. In terms of the ingoing and the outgoing Green functions, these take the form
\begin{equation}\label{eq:sRsLSin}
    \begin{split}
        &s_{\sR}(\zeta,k)  \equiv  \ \left(1-n^{\fd}_{k^0}\right) \left[\Sin(\zeta,k) -\Sout(\zeta,k)\right] \ ,  \\
        &s_{\sL}(\zeta,k) \equiv  \ -n^{\fd}_{k^0} \Sin(\zeta,k)-  \left(1-n^{\fd}_{k^0}\right) \Sout(\zeta,k) \ .
    \end{split}
\end{equation}
The Dirac conjugate solutions can also be arranged in the same form as in Eq.\eqref{eq:FermionSonTeaney}. Thus, the conjugate solutions have a very similar structure as those given above. We refer the reader to Appendix~(\ref{app:ConjDirac}) for more details.

\subsection{Yukawa theory}

We now have all the basic setup and notation we require to discuss Yukawa theory in the grSK spacetime. The action we consider is the sum of the free actions we have described in the previous subsections as well as the Yukawa interaction term:
\begin{equation}
    S = S_{\phi} + S_{\Psi} - i\oint \diff \zeta \int \diff^d x  \sqrt{-g} \ \lambda \phi \bar{\Psi} \Psi \ .
    \label{eq:ActionYukawa}
\end{equation}

Varying the above action gives Klein-Gordon and Dirac equations that are sourced by terms linear in the Yukawa coupling:
\begin{equation}
    \frac{1}{\sqrt{-g}} \partial_{\B} \left( \sqrt{-g} g^{\A \B} \partial_{\A}\phi\right)  = i\lambda \bar{\Psi} \Psi \  \qquad \text{and} \qquad \Gamma^{\A} D_{\A} \Psi = \lambda \phi \Psi\ .
\end{equation}
These equations can be solved perturbatively in the Yukawa coupling $\lambda$. In other words, we can express the solutions as
\begin{equation}
    \Psi = \Psi_{(0)} + \lambda \Psi_{(1)} + \lambda^2 \Psi_{(2)} + \ldots \  ,
    \label{eq:FormalPertExpPsi}
\end{equation}
\begin{equation}
    \phi = \phi_{(0)} + \lambda \phi_{(1)} + \lambda^2 \phi_{(2)} + \ldots \ .
    \label{eq:FormalPertExpphi}
\end{equation}
The leading-order solutions $\Psi_{(0)}$ and $\phi_{(0)}$ are just the free theory solutions that we have seen in detail in the previous subsections. We will take these solutions to satisfy the GKPW  boundary conditions we require. Thus the higher-order terms in the solutions will be normalisable on both the boundaries. To obtain the higher-order solutions, we will now need to develop a perturbation theory to solve sourced Klein-Gordon and Dirac equations. 

We will solve this problem using the Green function technique. First we define the binormalisable bulk-to-bulk Green functions. These are objects that are normalisable on both the boundaries of the grSK spacetime and satisfy the equations of motion with a Dirac-delta source in the bulk. More precisely, the scalar binormalisable Green function $\bbG(\zeta|\zeta_0,k)$ satisfies
\begin{equation}
   \left( -\Box +m^2 \right) \bbG(\zeta|\zeta_0,k) = \frac{ \delta(\zeta - \zeta_0)}{\sqrt{-g}} \ ,
\end{equation}
where $\delta(\zeta-\zeta_0)$ denotes the radial delta function on the grSK contour. The above differential equation for the massless case, can be explicitly written as
\begin{equation}
     \mathbb{D}_+ \left(r^{d-1} \mathbb{D}_+ \bbG(\zeta|\zeta_0,k)\right) + \frac{\beta^2}{4}r^{d-1} \left(f |\bk|^2 -(k^0)^2\right)\bbG(\zeta|\zeta_0,k) = \red{-} \frac{i \beta}{2} \delta(\zeta - \zeta_0)\ ,
     \label{eq:DESpinorBlkBlk}
\end{equation}
supplemented with the following boundary conditions
\begin{equation}
    \lim_{\zeta \to 0} \mathbb{G}(\zeta|\zeta_0,k) =\lim_{\zeta \to 1} \mathbb{G}(\zeta|\zeta_0,k) = 0 \ .
    \label{eq:BCSpinorBlkBlk}
\end{equation}

Similarly, the spinor bulk-to-bulk Green function $\bbS(\zeta|\zeta_0,k)$ is a square matrix in the bulk-spinor space that satisfies
\begin{equation}
    (-\Gamma^{\A} D_{\A}+m)\bbS(X|X')= \frac{\delta^{D}(X-X')}{\sqrt{-g}} \mathbb{1} \ .
\end{equation}
Written explicitly for the massless case, we have
\begin{equation}\label{eq:bbSODE}
    \begin{split}
        \left\{ \frac{\bbT(\zeta)}{\sqrt{f}}\partial _\zeta + \mathbb{\Gamma} \left( \beta k^0 + \frac{1}{2} \partial_\zeta \ln f \right) - \frac{\beta}{2} \Gamma^{(i)} k_i  \right\}\left(r^{d/2} \bbS (\zeta|\zeta_0,k) \right)  = \red{-} \mathbb{1} \frac{\delta(\zeta-\zeta_0)}{r^{d/2}f} \ ,
    \end{split}
\end{equation}
where $\mathbb{1}$ is the identity in the bulk spinor space, along with the boundary conditions
\begin{equation}\label{eq:BCbbS}
    \lim_{\zeta \to 0}r^{\frac{d}{2}} \Pplus S_{_{0}}\bbS(\zeta|\zeta_0,k) = 0 \ , \qquad \qquad \lim_{\zeta \to 1}r^{\frac{d}{2}} \Pplus S_{_{0}}\bbS(\zeta|\zeta_0,k) = 0 \ .
\end{equation}


An explicit derivation of the scalar bulk-to-bulk Green function can be found in \cite{Loganayagam:2024mnj}. As for the spinor bulk-to-bulk Green function, explicit expressions in terms of the boundary-to-bulk Green functions can be found in Appendix (\ref{sec:DerivationSpinorialBlkBlk}), where we derive the Green function along similar lines as in \cite{Faulkner:2013bna}, using the technique of Wronskians.

With the bulk-to-bulk Green functions in our hands, we can immediately write down expressions for the higher-order terms in the perturbative expansion of the solutions. For all $n>0$, we have 
\begin{equation}
    \Psi_{(n)} (\zeta,k) = \oint_{\zeta} \ \bbS(\zeta|\zeta_0,k)  \mathfrak{J}_{(n)} (\zeta_0,k)\ , \quad \text{and} \quad \phi_{(n)} (\zeta,k) = \oint_{\zeta} \ \bbG(\zeta|\zeta_0,k)  \mathbb{J}_{(n)} (\zeta_0,k) \ ,
    \label{eq:Psinandphin}
\end{equation}
where $\mathfrak{J}_{(n)} (\zeta_0,k)$ and $\mathbb{J}_{(n)} (\zeta_0,k)$ are sources for the $n$th order term in the solutions. These can be read off from the equations of motion. For example, at first order we have
\begin{equation}
    \begin{split}
        \mathfrak{J}_{(1)} (\zeta_0,k) &\equiv  \red{-} \int_{k_1} \int_{k_2} (2\pi)^{d} \delta^{(d)} (k_1+k_2-k) \phi_{(0)}(\zeta_0,k_1) \Psi_{(0)}(\zeta_0,k_2) \ , \\
        \mathbb{J}_{(1)} (\zeta_0,k) &\equiv  \red{-} i\int_{k_1} \int_{k_2} (2\pi)^{d} \delta^{(d)} (k_1+k_2-k) \overline{\Psi}_{(0)}(\zeta_0,k_1) \Psi_{(0)}(\zeta_0,k_2) \ .
    \end{split}
    \label{eq:BulkSources1}
\end{equation}
Similarly, at second-order,
\begin{equation}
    \begin{split}
        \mathfrak{J}_{(2)} (\zeta_0,k) &\equiv  \red{-} \int_{k_1} \int_{k_2} (2\pi)^{d} \delta^{(d)} (k_1+k_2-k) \left[ \phi_{(0)}(\zeta_0,k_1) \Psi_{(1)}(\zeta_0,k_2)+ \phi_{(1)}(\zeta_0,k_1) \Psi_{(0)}(\zeta_0,k_2) \right] \ ,\\
        \mathbb{J}_{(2)} (\zeta_0,k) &\equiv  \red{-} i\int_{k_1} \int_{k_2} (2\pi)^{d} \delta^{(d)} (k_1+k_2-k) \left[\overline{\Psi}_{(0)}(\zeta_0,k_1) \Psi_{(1)}(\zeta_0,k_2)+\overline{\Psi}_{(1)}(\zeta_0,k_1) \Psi_{(0)}(\zeta_0,k_2) \right] \ .
    \end{split}
    \label{eq:BulkSources2}
\end{equation}

\section{Yukawa theory and exterior EFT from grSK}\label{sec:YukawaandExteriorfromgrSK}

We now come to the influence phase for Yukawa theory. In the previous section, we have detailed how to obtain the solutions to the bulk equations of motion to arbitrary orders in the Yukawa coupling constant. Substituting these solutions into the action will then give us the tree-level influence phase up to any desired order. 

We will arrange this expansion with the notation
\begin{equation}
    S_{\text{on-shell}} = S_{(2)} + \lambda S_{(3)} + \lambda^2 S_{(4)} + \lambda^3S_{(5)} + \ldots \ ,
    \label{eq:OnShellActionPertExp}
\end{equation}
where $S_{\text{on-shell}}$ is the full on-shell action. On the RHS, we have employed a notation wherein the subscripts indicate the number of boundary sources in the term. Thus, for example, $S_{(2)}$ is quadratic in the boundary sources and takes the form given in Eq.~\eqref{eq:ActionS2}.

The term linear in the coupling $\lambda$ is also cubic in the boundary sources and takes the form
\begin{equation}
    S_{(3)} = -i \int_{k_{1,2,3}} \oint_{\zeta} \Psibar_{(0)}(\zeta,k_1)    \phi_{(0)}(\zeta,k_2) \Psi_{(0)}(\zeta,k_3) \ .
    \label{eq:ThreePtInfPhaseinSonTeaney}
\end{equation}
An explicit derivation of this can be found in Appendix \ref{sec:AppendixPertExpInfPhase}.
Here we have introduced a useful notation for the boundary momentum integrals, that we will employ throughout the rest of this note, 
\begin{equation}
    \int_{k_{1,2, \ldots, n}} \equiv \int \frac{\diff^d k_1}{(2\pi)^d} \int \frac{\diff^d k_2}{(2\pi)^d} \ldots \int \frac{\diff^d k_n}{(2\pi)^d}(2\pi)^d \delta^d(k_1+k_2+\ldots + k_n) \ .
    \label{eq:DefMomentumIntegral}
\end{equation}

The reader will notice that the three-point influence phase can be directly obtained from the free solutions in the usual Witten-diagram manner with the diagrams drawn on the full grSK contour. We will soon perform the monodromy integrals and see how this answer can be arranged in terms of Witten diagrams on one copy of the spacetime (instead of two, as in the grSK prescription).

The next correction, i.e., the term quadratic in the interaction parameter $\lambda$, which is quartic in the boundary sources, takes the form
\begin{equation}
    \begin{split}
        S_{(4)} =  \  &  \red{-} \frac{1}{2} \int_{k_{1,2,3,4}} \oint_{\z_1} \oint_{\z_2}  \overline{\Psi}_{(0)}(\z_2,k_1) \Psi_{(0)}(\z_2,k_2) \bbG(\z_2 |\z_1,k_3 +k_4) \overline{\Psi}_{(0)}(\z_1,k_3) \Psi_{(0)}(\z_1,k_4)\\
        & \red{-} -i \int_{k_{1,2,3,4}} \oint_{\z_1} \oint_{\z_2} \phi_{(0)}(\z_2,k_1)  \overline{\Psi}_{(0)}(\z_2,k_2)  \bbS(\z_2 |\z_1,k_3 +k_4) \phi_{(0)}(\z_1,k_3) \Psi_{(0)}(\z_1,k_4) \ .
    \end{split}
    \label{eq:FourPtInfPhaseinSonTeaneySols}
\end{equation}
Once again, an explicit derivation of the above results can be found in Appendix \ref{sec:AppendixPertExpInfPhase}.


We have seen above the three-point and the four-point influence phase written as integrals over the full grSK contour. We will now perform these monodromy integrals. To this end, let us focus on one particular term in the three-point influence phase. It will be clear from this discussion how the other terms can be similarly computed. We focus on the term with two future sources and one past source, each of which can be a scalar, a spinor, or its conjugate. Thus we expect three terms at the end. Starting from Eq.~\eqref{eq:ThreePtInfPhaseinSonTeaney} and extracting the desired term, we have
\begin{equation}
    \begin{split}
        S_{\Fb \Fb \Pb } &= -i \int_{k_{1,2,3}} \oint_{\z} \Bigg\{  \left[-\psibF(k_1) \Sbin (\z,k_1)\right] \left[ - \Gin(\z, k_2) J_{\Fb}(k_2)\right]\left[-e^{\beta k^0_3}\Sout(\z, k_3) \psiP(k_3)\right]\\
        &\hspace{2cm}+\left[-\psibF(k_1) \Sbin (\z,k_1)\right] \left[ e^{\beta k^0_2}\Gout(\z, k_2) J_{\Pb}(k_2)\right]\left[-\Sin(\z, k_3) \psiF(k_3)\right]\\
        &\hspace{2cm}+\left[-e^{\beta k^0_1}\psibP(k_1) \Sbout (\z,k_1)\right] \left[ - \Gin(\z, k_2) J_{\Fb}(k_2)\right]\left[-\Sin(\z, k_3) \psiF(k_3)\right]\Bigg\} \ .
    \end{split}
\end{equation}
The ingoing Green function is analytic on the grSK contour and simply goes for a ride in the monodromy integral. The only non-analyticities in the above expression come from the outgoing Green functions. These come from the factors of $\sqrt{f}$ and $e^{-\beta k^0 \zeta}$. Noting that $\sqrt{f}$ flips in sign while going from one branch of the contour to the other, and $\zeta$ jumps by unity, we get
\begin{equation}
    \oint_{\zeta} e^{-\beta k^0 \zeta} f^{j/2} \mathcal{F}(\zeta) = \int_{\rm ext} e^{-\beta k^0 \zeta} f^{j/2} \mathcal{F}(\zeta) \times
    \begin{cases}
        \frac{-1}{1+n^{\be}_{k^0}} & \text{for} \ j=0\\
        \frac{-1}{1-n^{\fd}_{k^0}} & \text{for} \  j=1
    \end{cases}\ ,
\end{equation}
where $\mathcal{F}(\z)$ is an analytic function and we have chosen to write the integral over the left branch of the grSK contour with $\int_{\rm ext}$ defined as
\begin{equation}
    \int_{\rm ext} \equiv \int_{r_h}^{r_c} \diff r \ r^{d-1}\ .
    \label{eq:DefExteriorIntegral}
\end{equation}
Using this result, we obtain
\begin{equation}\label{eq:3ptFFP}
    \begin{split}
        S_{\Fb \Fb \Pb} &= -i \int_{k_{1,2,3}} \int_{\rm ext} \bigg\{\frac{1}{n^{\fd}_{k_3^0}} \left[\psibF(k_1) \Sbin (\z,k_1)\right] \left[\Gin(\z, k_2) J_{\Fb}(k_2)\right]\left[\Sout(\z, k_3) \psiP(k_3)\right]\\
        &\hspace{2.8cm}-\frac{1}{n^{\be}_{k_2^0}}\left[\psibF(k_1) \Sbin (\z,k_1)\right] \left[\Gout(\z, k_2) J_{\Pb}(k_2)\right]\left[\Sin(\z, k_3) \psiF(k_3)\right]\\
        &\hspace{2.8cm}+\frac{1}{n^{\fd}_{k_1^0}}\left[\psibP(k_1) \Sbout (\z,k_1)\right] \left[  \Gin(\z, k_2) J_{\Fb}(k_2)\right]\left[\Sin(\z, k_3) \psiF(k_3)\right]\bigg\} \ ,
    \end{split}
\end{equation}
Note also that none of the functions in the integrand have poles at the horizon (which can be gleaned from the explicit expressions of the Green functions in \cite{Jana:2020vyx, Loganayagam:2020eue}), and thus there are no horizon-localised contributions.

We can now perform similar computations for all the other terms in the three-point influence phase. The term with only future sources vanishes, i.e., $S_{\Fb \Fb \Fb} = 0$, which is simply the SK collapse condition at the three-point level. This is because the integrand is completely analytic and thus the monodromy integral vanishes. Similarly, the term with all past sources also vanishes, i.e., $S_{\Pb \Pb \Pb} = 0$, which is the KMS condition at the three-point level. This is because each term has an exponential factor, all of which multiply to give unity by momentum conservation, and a factor of $f$ which is analytic around $r=r_h$. Therefore, the only other non-vanishing term in the three-point influence phase is the term with two past sources. This term looks like

\begin{equation}\label{eq:3ptFPP}
    \begin{split}
        S_{\Fb \Pb \Pb} &= -i \int_{k_{1,2,3}} \int_{\rm ext} \bigg\{  -\frac{1}{n^{\fd}_{k_2^0+k_3^0}} \left[\psibF(k_1) \Sbin (\z,k_1)\right] \left[  \Gout(\z, k_2) J_{\Pb}(k_2)\right]\left[\Sout(\z, k_3) \psiP(k_3)\right]\\
        &\hspace{3cm}+\frac{1}{n^{\be}_{k_1^0+k_3^0}} \left[\psibP(k_1) \Sbout (\z,k_1)\right] \left[  \Gin(\z, k_2) J_{\Fb}(k_2)\right]\left[\Sout(\z, k_3) \psiP(k_3)\right]\\
        &\hspace{3cm}-\frac{1}{n^{\fd}_{k_1^0+k_2^0}}\left[\psibP(k_1) \Sbout (\z,k_1)\right] \left[\Gout(\z, k_2) J_{\Pb}(k_2)\right]\left[\Sin(\z, k_3) \psiF(k_3)\right] \bigg\} \ .
    \end{split}
\end{equation}
Thus the full three-point influence phase takes the form
\begin{equation}
    S_{(3)} = S_{\Fb \Fb \Pb} + S_{\Fb \Pb \Pb} \ .
\end{equation}

We can compute the higher point contributions to the influence phase using similar arguments as above. Rather than illustrating explicit calculations, we will introduce a series of diagrammatic Feynman rules. Through these rules, the reader can determine the on-shell action at any desired order.

\subsection{Feynman rules for Yukawa Theory}

The diagrammatic rules for the computation of the tree-level influence phase (on-shell action) are given below. The convention we use for the direction of momentum flow is as follows. In any boundary-to-bulk propagator, the momentum flows from the boundary to the bulk. In the bulk-to-bulk propagators, the momentum flows from the left to the right.

The scalar boundary-to-bulk propagators are as follows.
\begin{equation}
    \begin{tikzpicture}[scale=1.5]
        \draw[blue] (-0.5,0)--(0.5,0);  
        \Gdiode{0}{0}{0}{-1.7};
        \node at (0,-1.7) {$\bullet$};
        \node at (0.2,-1.7) {$\zeta$};
        \node at (-0.3,-0.5) {$k$};
        \node at (1.65, -0.75) {$\equiv \frac{\Gin(\zeta,k)}{1+n^{\be}_{k^0}}\ J_{\Fb}(k)\ ,$};
    \end{tikzpicture}
    \hspace{2cm}
    \begin{tikzpicture}[scale =1.5]
        \draw[blue] (-0.5,0)--(0.5,0);  
        \Gdiode{0}{-1.7}{0}{0};
        \node at (0,-1.7) {$\bullet$};
        \node at (0.2,-1.7) {$\zeta$};
        \node at (-0.3,-0.5) {$k$};
        \node at (2.5, -0.75) {$\equiv  G^{\rm out}(\zeta,k )\ J_{\Pb}(k)  \ .$};
    \end{tikzpicture}
\end{equation}
The scalar bulk-to-bulk propagators are
\begin{equation}
    \begin{tikzpicture}[scale=1.5]
        \Gdiode{-1}{0}{1}{0};
        \node at (-0.3,-0.2) {$k$};
        \node at (-1,-0.2) {$\zeta_1$};
        \node at (1,-0.2) {$\zeta_2$};
        \node at (-1,0) {$\bullet$};
        \node at (1,0) {$\bullet$};
        \node at (2.5,0) {$ \equiv  -i \bbGR (\zeta_2|\zeta_1,k) \ ,$};
    \end{tikzpicture}
    \hspace{2cm}
    \begin{tikzpicture}[scale=1.5]
        \Gdiode{1}{0}{-1}{0};
        \node at (-0.3,-0.2) {$k$};
        \node at (-1,-0.2) {$\zeta_1$};
        \node at (1,-0.2) {$\zeta_2$};
        \node at (-1,0) {$\bullet$};
        \node at (1,0) {$\bullet$};
        \node at (2.5,0) {$ \equiv  -i \bbGA (\zeta_2|\zeta_1,k) \ .$};
    \end{tikzpicture}
\end{equation}

The spinor boundary-to-bulk propagators are given below. Here the additional arrows represent the spinorial direction ($\psibar$ to $\psi$), which becomes the charge arrow in QED. 
\begin{equation}
    \begin{tikzpicture}[scale=1.5]
        \draw[blue] (-0.5,0)--(0.5,0);  
        \Sdiode{0}{0}{0}{-1.7};
        \node at (0,-1.7) {$\bullet$};
        \node at (0.2,-1.7) {$\zeta$};
        \node at (0.3,-0.5) {$k$};
        \node at (-0.3,-0.7) {$\uparrow$};
        \node at (1.65, -0.75) {$\equiv  \frac{\Sin(\zeta,k)}{1-n^{\fd}_{k^0}} \ \psi_{\Fb}(k)\ ,$};
    \end{tikzpicture}
    \hspace{2cm}
    \begin{tikzpicture}[scale =1.5]
        \draw[blue] (-0.5,0)--(0.5,0);  
        \Sdiode{0}{-1.7}{0}{0};
        \node at (0,-1.7) {$\bullet$};
        \node at (0.2,-1.7) {$\zeta$};
        \node at (0.3,-0.5) {$k$};
        \node at (-0.3,-0.7) {$\uparrow$};
        \node at (2.5, -0.75) {$\equiv  \Sout(\zeta,k ) \ \psi_{\Pb}(k) \ ,$};
    \end{tikzpicture}
\end{equation}
\begin{equation}
    \begin{tikzpicture}[scale=1.5]
        \draw[blue] (-0.5,0)--(0.5,0);  
        \Sdiode{0}{0}{0}{-1.7};
        \node at (0,-1.7) {$\bullet$};
        \node at (0.2,-1.7) {$\zeta$};
        \node at (0.3,-0.5) {$k$};
        \node at (-0.3,-0.7) {$\downarrow$};
        \node at (1.65, -0.75) {$\equiv  \frac{\Sbin(\zeta,k)}{1-n^{\fd}_{k^0}} \ i\psibar_{\Fb}(k)\ ,$};
    \end{tikzpicture}
    \hspace{2cm}
    \begin{tikzpicture}[scale =1.5]
        \draw[blue] (-0.5,0)--(0.5,0);  
        \Sdiode{0}{-1.7}{0}{0};
        \node at (0,-1.7) {$\bullet$};
        \node at (0.2,-1.7) {$\zeta$};
        \node at (0.3,-0.5) {$k$};
        \node at (-0.3,-0.7) {$\downarrow$};
        \node at (2.5, -0.75) {$\equiv  \Sbout(\zeta,k ) \ i\psibar_{\Pb}(k) \ .$};
    \end{tikzpicture}
\end{equation}
The spinor bulk-to-bulk propagators are
\begin{equation}
    \begin{tikzpicture}[scale=1.5]
        \Sdiode{-1}{0}{1}{0};
        \node at (-0.3,-0.2) {$\rightarrow$};
        \node at (-0.3,0.2) {$k$};
        \node at (-1,-0.2) {$\zeta_1$};
        \node at (1,-0.2) {$\zeta_2$};
        \node at (-1,0) {$\bullet$};
        \node at (1,0) {$\bullet$};
        \node at (2.5,0) {$ \equiv  -i \bbSR (\zeta_2|\zeta_1,k) \ ,$};
    \end{tikzpicture}
    \hspace{2cm}
    \begin{tikzpicture}[scale=1.5]
        \Sdiode{1}{0}{-1}{0};
        \node at (-0.3,-0.2) {$\rightarrow$};
        \node at (-0.3,0.2) {$k$};
        \node at (-1,-0.2) {$\zeta_1$};
        \node at (1,-0.2) {$\zeta_2$};
        \node at (-1,0) {$\bullet$};
        \node at (1,0) {$\bullet$};
        \node at (2.5,0) {$ \equiv  -i \bbSA (\zeta_2|\zeta_1,k) \ .$};
    \end{tikzpicture}
\end{equation}

Finally, the vertices are as follows.
\begin{equation}
    \begin{tikzpicture}[scale=1.1]
        \Sdiodearrow{0}{0}{0}{1}
        \node at (0,0) {$\bullet$};
        \node at (0.4,0.9) {$k_1$};
        \Sdiodearrow{0}{0}{-1}{-0.5}
        \node at (-1.1,0) {$k_2$};
        \Gsemicap{1}{-0.5}{0}{0}
        \node at (1,0) {$k_3$};
        \node at (4,0) {$=  \red{i} \lambda  \frac{n^{\be}_{-k^0_3}}{n^{\be}_{k^0_1+k^0_2}} \ ,$};
    \end{tikzpicture}
    \qquad
    \begin{tikzpicture}[scale=1.1]
        \Sdiodearrow{0}{0}{0}{1}
        \node at (0,0) {$\bullet$};
        \node at (0.4,0.9) {$k_1$};
        \Ssemicap{-1}{-0.5}{0}{0}
        \node at (-1,0) {$k_2$};
        \Gsemicap{1}{-0.5}{0}{0}
        \node at (1,0) {$k_3$};
        \node at (4,0) {$= \red{i} \lambda  \frac{-n^{\fd}_{-k^0_2} n^{\be}_{-k^0_3}}{-n^{\fd}_{k^0_1}} \ ,$};
    \end{tikzpicture}
\end{equation}
\begin{equation}
    \begin{tikzpicture}[scale=1.1]
        \Gdiodearrow{0}{0}{0}{1}
        \node at (0,0) {$\bullet$};
        \node at (0.5,0.8) {$k_1$};
        \Sdiodearrow{0}{0}{-1}{-0.5}
        \node at (-1.1,0) {$k_2$};
        \Ssemicap{1}{-0.5}{0}{0}
        \node at (1,0) {$k_3$};
        \node at (4,0) {$=  \red{i} \lambda  \frac{-n^{\fd}_{-k^0_3}}{-n^{\fd}_{k^0_1+k^0_2}} \ ,$};
    \end{tikzpicture}
    \qquad
    \begin{tikzpicture}[scale=1.1]
        \Gdiodearrow{0}{0}{0}{1}
        \node at (0,0) {$\bullet$};
        \node at (0.5,0.8) {$k_1$};
        \Ssemicap{-1}{-0.5}{0}{0}
        \node at (-1,0) {$k_2$};
        \Ssemicap{1}{-0.5}{0}{0}
        \node at (1,0) {$k_3$};
        \node at (4,0) {$= \red{i} \lambda  \frac{(-n^{\fd}_{-k^0_2})(-n^{\fd}_{-k^0_3})}{n^{\be}_{k^0_1}} \ .$};
    \end{tikzpicture}
\end{equation}
In each of the above vertices, we have adopted the convention that all the momenta flow into the vertex. The reader may note that these rules are the same as those given in \cite{Gelis:2019yfm}.

Along with the above Feynman rules, to obtain the influence phase, we must further 
\begin{enumerate}
    \item  Multiply every diagram by $\red{-}i$.
    \item The vertices are integrated over the exterior of the blackbrane, with the following  radial exterior integral
\begin{equation}\label{eq:extint}
    \int_{\rm ext} = \int_{r_{h}}^{r_{c}}\diff r \, r^{d-1}  \ .
\end{equation}
    \item  Impose momentum conservation at each vertex and integrate over all momenta.
    \item Divide by the symmetry factor of each diagram.
\end{enumerate}


Notice that in the vertices above, the statistical factors in the denominator are functions of the sum of the momenta running over legs ending in a triangle. When the number of such fermionic legs is even, we get a Bose-Einstein factor, and when it is odd, we get a Fermi-Dirac factor. This seems to be a general feature in our calculations. When statistical factors are functions of sum over momenta of two (more generally even) fermionic legs, one effectively gets a bosonic statistical factor.

\begin{figure}[H]
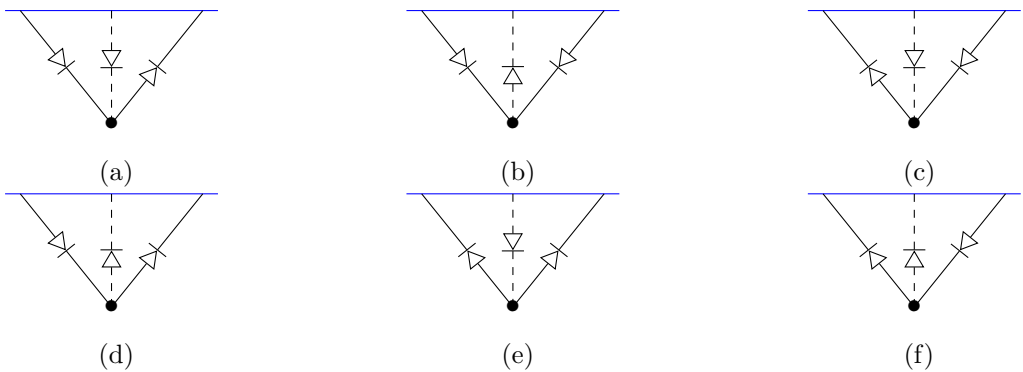

    \centering
    \begin{subfigure}{0.33\textwidth}
        \centering
        \threeptPFYuk{1}{1}{-1}
        \caption{}
    \end{subfigure}%
    \begin{subfigure}{0.33\textwidth}
        \centering
        \threeptPFYuk{1}{-1}{1}
        \caption{}
    \end{subfigure}%
    \begin{subfigure}{0.33\textwidth}
        \centering
        \threeptPFYuk{-1}{1}{1}
        \caption{}
    \end{subfigure}\\
    \begin{subfigure}{0.33\textwidth}
        \centering
        \threeptPFYuk{1}{-1}{-1}
        \caption{}
    \end{subfigure}%
    \begin{subfigure}{0.33\textwidth}
        \centering
        \threeptPFYuk{-1}{1}{-1}
        \caption{}
    \end{subfigure}%
    \begin{subfigure}{0.33\textwidth}
        \centering
        \threeptPFYuk{-1}{-1}{1}
        \caption{}
    \end{subfigure}\\
    \caption{Witten diagrams that contribute to the three-point influence phase.}
    \label{fig:ThreePointYukawa}
\end{figure}

With the rules above, the diagrams that contribute to the three-point influence phase are given in Fig.~(\ref{fig:ThreePointYukawa}). The reader can easily check that the expressions in Eq.~\eqref{eq:3ptFPP} and Eq.~\eqref{eq:3ptFFP} are consistent with those generated by these diagrams with the rules above. In all the diagrams below, the blue line denotes the left AdS boundary. Here, the charge arrow of the spinors are always taken to be flowing from the boundary to the bulk, and from the left to the right in the bulk-to-bulk propagators.

Similarly, the diagrams that contribute to the four-point influence phase are as follows.
\begin{figure}[H]
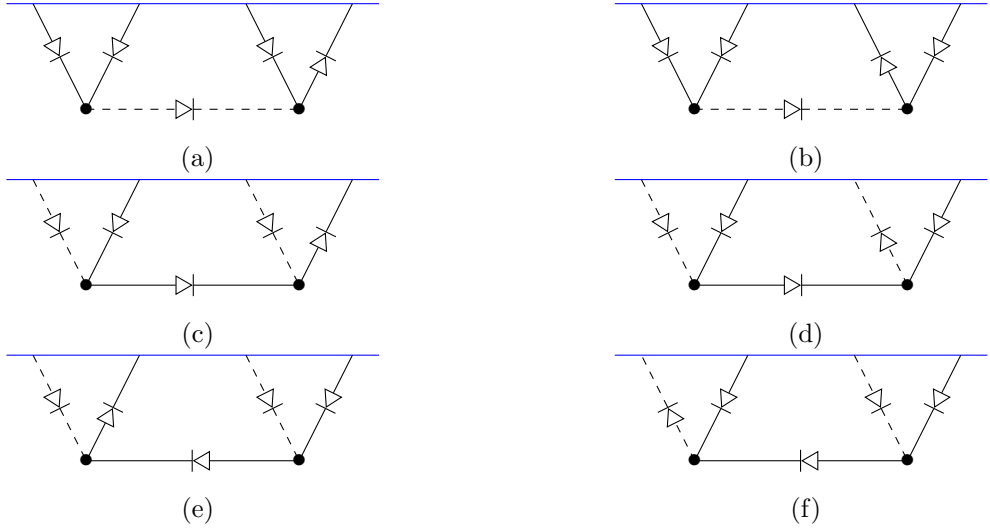

    \centering
    \begin{subfigure}{0.5\textwidth}
        \centering
        \fourptScalExPFYuk{1}{1}{1}{-1}{1}
        \caption{}
    \end{subfigure}%
    \begin{subfigure}{0.5\textwidth}
        \centering
        \fourptScalExPFYuk{1}{1}{-1}{1}{1}
        \caption{}
    \end{subfigure}\\
    \begin{subfigure}{0.5\textwidth}
        \centering
        \fourptFermExPFYuk{1}{1}{1}{-1}{1}
        \caption{}
    \end{subfigure}%
    \begin{subfigure}{0.5\textwidth}
        \centering
        \fourptFermExPFYuk{1}{1}{-1}{1}{1}
        \caption{}
    \end{subfigure}\\
    \begin{subfigure}{0.5\textwidth}
        \centering
        \fourptFermExPFYuk{1}{-1}{1}{1}{-1}
        \caption{}
    \end{subfigure}%
    \begin{subfigure}{0.5\textwidth}
        \centering
        \fourptFermExPFYuk{-1}{1}{1}{1}{-1}
        \caption{}
    \end{subfigure}\\
    \caption{Witten diagrams that contribute to the $S_{\Fb \Fb \Fb \Pb}$ term in the four-point exchange influence phase.}
    \label{fig:FFFPExchange}
\end{figure}

\begin{figure}[H]
    \centering
    \begin{subfigure}{0.5\textwidth}
        \centering
        \fourptScalExPFYuk{1}{-1}{-1}{-1}{1}
        \caption{}
    \end{subfigure}%
    \begin{subfigure}{0.5\textwidth}
        \centering
        \fourptScalExPFYuk{-1}{1}{-1}{-1}{1}
        \caption{}
    \end{subfigure}\\
    \begin{subfigure}{0.5\textwidth}
        \centering
        \fourptFermExPFYuk{1}{-1}{-1}{-1}{1}
        \caption{}
    \end{subfigure}%
    \begin{subfigure}{0.5\textwidth}
        \centering
        \fourptFermExPFYuk{-1}{1}{-1}{-1}{1}
        \caption{}
    \end{subfigure}\\
    \begin{subfigure}{0.5\textwidth}
        \centering
        \fourptFermExPFYuk{-1}{-1}{1}{-1}{-1}
        \caption{}
    \end{subfigure}%
    \begin{subfigure}{0.5\textwidth}
        \centering
        \fourptFermExPFYuk{-1}{-1}{-1}{1}{-1}
        \caption{}
    \end{subfigure}\\
    \caption{Witten diagrams that contribute to the $S_{\Fb \Pb \Pb \Pb}$ term in the four-point exchange influence phase.}
    \label{fig:FPPPExchange}
\end{figure}

\begin{figure}[H]
    \centering
    \begin{subfigure}{\textwidth}
        \centering
        \fourptScalExPFYuk{1}{1}{-1}{-1}{1}
        \caption{}
    \end{subfigure}\\
    \begin{subfigure}{0.5\textwidth}
        \centering
        \fourptFermExPFYuk{1}{1}{-1}{-1}{1}
        \caption{}
    \end{subfigure}%
    \begin{subfigure}{0.5\textwidth}
        \centering
        \fourptFermExPFYuk{-1}{-1}{1}{1}{-1}
        \caption{}
    \end{subfigure}\\
    \caption{Witten diagrams that contribute to the $S_{\Fb \Fb \Pb \Pb}$ term in the four-point exchange influence phase.}
    \label{fig:FFPPExchange}
\end{figure}

\begin{figure}[H]
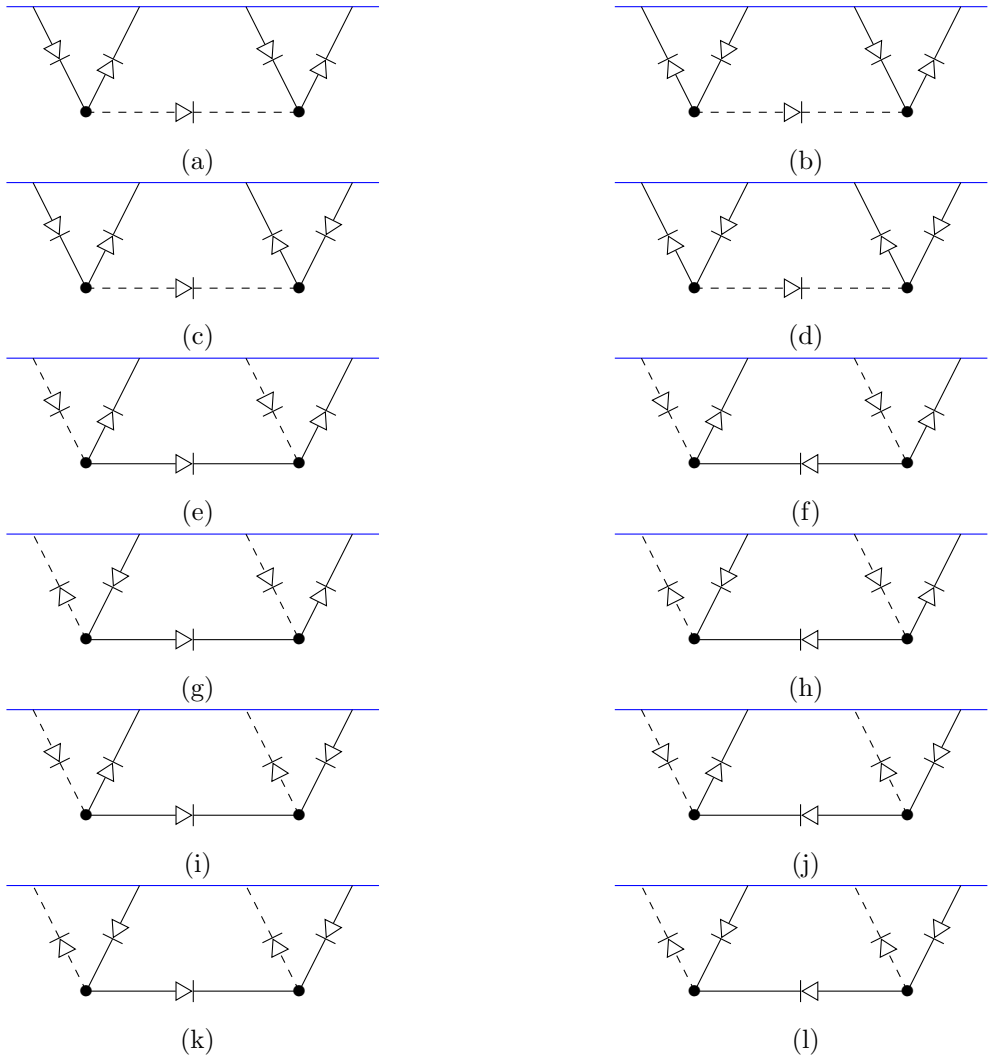

    \centering
    \begin{subfigure}{0.5\textwidth}
        \centering
        \fourptScalExPFYuk{1}{-1}{1}{-1}{1}
        \caption{}
    \end{subfigure}%
    \begin{subfigure}{0.5\textwidth}
        \centering
        \fourptScalExPFYuk{-1}{1}{1}{-1}{1}
        \caption{}
    \end{subfigure}\\
    \begin{subfigure}{0.5\textwidth}
        \centering
        \fourptScalExPFYuk{1}{-1}{-1}{1}{1}
        \caption{}
    \end{subfigure}%
    \begin{subfigure}{0.5\textwidth}
        \centering
        \fourptScalExPFYuk{-1}{1}{-1}{1}{1}
        \caption{}
    \end{subfigure}\\
    \begin{subfigure}{0.5\textwidth}
        \centering
        \fourptFermExPFYuk{1}{-1}{1}{-1}{1}
        \caption{}
    \end{subfigure}%
    \begin{subfigure}{0.5\textwidth}
        \centering
        \fourptFermExPFYuk{1}{-1}{1}{-1}{-1}
        \caption{}
    \end{subfigure}\\
    \begin{subfigure}{0.5\textwidth}
        \centering
        \fourptFermExPFYuk{-1}{1}{1}{-1}{1}
        \caption{}
    \end{subfigure}%
    \begin{subfigure}{0.5\textwidth}
        \centering
        \fourptFermExPFYuk{-1}{1}{1}{-1}{-1}
        \caption{}
    \end{subfigure}\\
    \begin{subfigure}{0.5\textwidth}
        \centering
        \fourptFermExPFYuk{1}{-1}{-1}{1}{1}
        \caption{}
    \end{subfigure}%
    \begin{subfigure}{0.5\textwidth}
        \centering
        \fourptFermExPFYuk{1}{-1}{-1}{1}{-1}
        \caption{}
    \end{subfigure}\\
    \begin{subfigure}{0.5\textwidth}
        \centering
        \fourptFermExPFYuk{-1}{1}{-1}{1}{1}
        \caption{}
    \end{subfigure}%
    \begin{subfigure}{0.5\textwidth}
        \centering
        \fourptFermExPFYuk{-1}{1}{-1}{1}{-1}
        \caption{}
    \end{subfigure}\\
    \caption{Witten diagrams that contribute to the $S_{\Fb \Pb \Fb \Pb}$ term in the four-point exchange influence phase.}
    \label{fig:FPFPExchange}
\end{figure}

\section{Column vector representation of Yukawa influence phase}\label{sec:YukawaIF}

In this section, we will express the three-point and the four-point Yukawa influence phase in a convenient notation, i.e., the \textit{column vector} representation. The column vector representation naturally gives the spectral functions in the real-time formalism \cite{Chaudhuri:2018ihk}. This is not surprising as real-time correlators admit K\"{a}ll\a'{e}n-Lehmann-like spectral representations \cite{Evans:1990hy,Evans:1991ky}. It has been very useful in perturbative techniques, computation of transport coefficients, etc. \cite{Henning:1993gh,Carrington:2006xj}. 

The column vector representation manifestly incorporates the microscopic unitarity and the thermality of the system as shown in \cite{Chaudhuri:2018ihk}. Here, the authors showed how thermal bosonic Out-of-Time-Ordered Correlators (OTOCs) can be organised in spectral representations using column vectors. While for scalar holographic theories, it has been shown that the same results can be obtained from the bulk \cite{Loganayagam:2024mnj}, the generalisation to include spinors has not been done so far. From a boundary perspective, we anticipate the existence of a similar representation for fermions. Here, we aim to demonstrate that this expectation is indeed supported by the bulk side as well.

We now turn our attention to the explicit construction of the column vector representation. In this representation, boundary scalar sources are expressed as
\begin{equation}
    \begin{split}
        &J_{\Fb}(k) \equiv \mathbf{J}(k) \cdot \eF^{{}_{(1)}}(\bar{k})  \ , \qquad  J_{\Pb}(k) \equiv \mathbf{J}(k) \cdot \eP^{{}_{(1)}}(\bar{k})\  ,
    \end{split}
\end{equation}
where we have employed the notation $\bar{k} \equiv -k$ to avoid clutter, and the symbol $\mathbf{J}(k)$ denotes the array of boundary scalar sources:
\begin{equation}
    \mathbf{J}(k) \equiv
    \begin{pmatrix}
        J_{\sR}(k) & \quad -J_{\sL}(k)
    \end{pmatrix} \ .
\end{equation}
Furthermore, $\eP(k)$ and $\eF(k)$ are column vectors of the form
\begin{equation}
    \eP^{{}_{(1)}}(k) \equiv 
    \begin{pmatrix}
        1+n^{\be}_{k^0}\\
        1+n^{\be}_{k^0}
    \end{pmatrix} \ ,\quad \text{and} \quad
    \eF^{{}_{(1)}}(k) \equiv 
    \begin{pmatrix}
         n^{\be}_{k^0}\\
         1+n^{\be}_{k^0}
    \end{pmatrix} \ .
\end{equation}

Similarly, we now introduce the column vector representation for spinors. In column vector notation, the boundary spinor sources are 
\begin{equation}
    \begin{split}
        &\psiF(k) \equiv \boldsymbol{\psi}(k) \cdot  \fF^{{}_{(1)}}(\bar{k})  \ , \qquad  \psiP(k) \equiv \boldsymbol{\psi}(k)\cdot  \fP^{{}_{(1)}}(\bar{k})  \ ,\\
        &\psibF(k) \equiv \boldsymbol{\psibar}(k) \cdot  \fF^{{}_{(1)}}(\bar{k})  \ , \qquad  \psibP(k) \equiv \boldsymbol{\psibar}(k)\cdot  \fP^{{}_{(1)}}(\bar{k})  \ ,
    \end{split}
\end{equation}
where $\boldsymbol{\psi}(k)$ and $\boldsymbol{\psibar}(k)$ are arrays of boundary spinor sources defined as
\begin{equation}
    \boldsymbol{\psi}(k) \equiv
    \begin{pmatrix}
        \psi_{\sR}(k) &\quad  -\psi_{\sL}(k)
    \end{pmatrix} \ , \qquad
    \boldsymbol{\psibar}(k) \equiv
    \begin{pmatrix}
        \psibar_{\sR}(k) & \quad -\psibar_{\sL}(k)
    \end{pmatrix} \ .
\end{equation}
Here again, $\fF^{{}_{(1)}}$ and $\fP^{{}_{(1)}}$ are two-dimensional column vectors whose explicit expressions are
\begin{equation}
    \fP^{{}_{(1)}}(k) \equiv  
    \begin{pmatrix}
        1-n^{\fd}_{k^0}\\
        1-n^{\fd}_{k^0}
    \end{pmatrix}
    \ , \quad 
    \fF^{{}_{(1)}}(k) \equiv
    \begin{pmatrix}
        -n^{\fd}_{k^0}\\
        1-n^{\fd}_{k^0}
    \end{pmatrix}
     \ .
\end{equation}
For convenience, we also define a second set of column vectors as 
\begin{equation}
    \begin{split}
        \eP^{{}_{(2)}}(k) &\equiv e^{-\beta k^0} \eP^{{}_{(1)}}(k) \ , \qquad \eF^{{}_{(2)}}(k) \equiv e^{-\beta k^0} \eF^{{}_{(1)}}(k) \ ,\\
        \fP^{{}_{(2)}}(k) &\equiv -e^{-\beta k^0} \fP^{{}_{(1)}}(k) \ , \qquad \fF^{{}_{(2)}}(k) \equiv -e^{-\beta k^0} \fF^{{}_{(1)}}(k) \ .
    \end{split}
\end{equation}
Note that the column vectors for spinors can be obtained from the scalar column vectors by replacing the Bose-Einstein factor $n^{\be}_{k^0}$ with the negative of the Fermi-Dirac factor $-n^{\fd}_{k^0}$, or equivalently every $e^{-\beta k^0}$ by $-e^{-\beta k^0}$. 

\subsection*{Three-point Yukawa Influence Phase}
We will now present the three-point Yukawa influence phase in Eq.~\eqref{eq:3ptFFP} and Eq.~\eqref{eq:3ptFPP}, in the column vector representation:
\begin{equation}
    \begin{split}
        &S_{\Fb \Fb \Pb} = i  \int_{k_{1,2,3}} \Bigg\{  \langle  \overline{S}^{\rm in} G^{\rm in} S^{\rm out} \rangle \cdot   \fF^{{}_{(1)}}(\bar{k}_1) \otimes \eF^{{}_{(1)}}(\bar{k}_2) \otimes \left[ \fP^{{}_{(2)}}(\bar{k}_3)-\fP^{{}_{(1)}}(\bar{k}_3)\right]\\
        &\hspace{2.5cm}+ \langle  \overline{S}^{\rm in} G^{\rm out} S^{\rm in} \rangle \cdot  \fF^{{}_{(1)}}(\bar{k}_1) \otimes \left[\eP^{{}_{(2)}}(\bar{k}_2)-\eP^{{}_{(1)}}(\bar{k}_2)\right] \otimes \fF^{{}_{(1)}}(\bar{k}_3) \\
        &\hspace{2.5cm}+\langle  \overline{S}^{\rm out} G^{\rm in} S^{\rm in} \rangle \cdot 
        \left[\fP^{{}_{(2)}}(\bar{k}_1)-\fP^{{}_{(1)}}(\bar{k}_1)\right] \otimes \eF^{{}_{(1)}}(\bar{k}_2) \otimes \fF^{{}_{(1)}}(\bar{k}_3) \Bigg\}\ ,
    \end{split}
\end{equation}
and
\begin{equation}
    \begin{split}
        &S_{\Fb \Pb \Pb} = -i  \int_{k_{1,2,3}} \Bigg\{ \langle \overline{S}^{\rm in} G^{\rm out} S^{\rm out} \rangle \cdot \fF^{{}_{(1)}}(\bar{k}_1) \otimes \left[ \eP^{{}_{(2)}}(\bar{k}_2) \otimes \fP^{{}_{(2)}}(\bar{k}_3)- \eP^{{}_{(1)}}(\bar{k}_2) \otimes \fP^{{}_{(1)}}(\bar{k}_3)\right] \\
        &\hspace{2.5cm}+\langle \overline{S}^{\rm out} G^{\rm in} S^{\rm out} \rangle \cdot \left[ \fP^{{}_{(2)}}(\bar{k}_1)  \otimes \fP^{{}_{(2)}}(\bar{k}_3)-\fP^{{}_{(1)}}(\bar{k}_1)  \otimes \fP^{{}_{(1)}}(\bar{k}_3)\right] \otimes \eF^{{}_{(1)}}(\bar{k}_2)\\
        &\hspace{2.5cm}+\langle \overline{S}^{\rm out} G^{\rm out} S^{\rm in} \rangle \cdot \left[\fP^{{}_{(2)}}(\bar{k}_1) \otimes \eP^{{}_{(2)}}(\bar{k}_2)-\fP^{{}_{(1)}}(\bar{k}_1) \otimes \eP^{{}_{(1)}}(\bar{k}_2)\right] \otimes \fF^{{}_{(1)}}(\bar{k}_3) \Bigg\} \ .
    \end{split}
\end{equation}
For notational convenience, we have used the definition\footnote{Here, $\bulkalpha,\bulkbeta,\ldots$ and $\alpha,\beta,\ldots$ denote the bulk and boundary spinor indices respectively.}
\begin{equation}
   \langle \overline{S}^{x_1} G^{x_2} S^{x_3} \rangle \equiv \int_{\rm ext} \left[\overline{S}^{x_1}(\zeta, k_1)\right]^{\alpha}_{\ \bulkalpha} G^{x_2}(\zeta, k_2) \left[S^{x_3}(\zeta, k_3)\right]^{\bulkalpha}_{\ \beta}  \left[ \boldsymbol{\psibar}(k_1)  \right]_{\alpha} \otimes \mathbf{J}(k_2) \otimes \left[\boldsymbol{\psi}(k_3)\right]^{\beta}\ , 
\end{equation}
where $x_1, x_2, x_3 \in \{\text{in}, \text{out}\}$. Notice that all the statistical factors are absorbed into the column vectors. We will see that something similar continues to happen even at four-point, thus proving the usefulness of the column vector representation. 

\subsection*{Four-point Yukawa Influence Phase}
Let's now come to the four-point Yukawa influence phase at the tree level. Here, we will find it convenient to classify all the terms and diagrams into two sets, one of which has a bulk scalar exchange and the other a bulk spinor exchange. Thus, we will write
\begin{equation}
    S_{(4)} = S_{4,\bbG} + S_{4,\bbS} \ ,
\end{equation}
where the $S_{4,\bbG}$ and the $S_{4,\bbS}$ denote the terms with the scalar and the spinor bulk-to-bulk propagators respectively. Furthermore, we will decompose each of these terms according to combinations of boundary sources in them. That is,
\begin{equation}
    \begin{split}
        S_{4,\bbG} &= S_{ \Fb \Fb \Fb \Pb,\bbG} + S_{\Fb \Fb \Pb \Pb,\bbG}+  S_{\Fb \Pb \Fb \Pb,\bbG}+ S_{\Fb \Pb \Pb \Pb,\bbG} \ ,  \\
        S_{4,\bbS} &= S_{ \Fb \Fb \Fb \Pb,\bbS} + S_{\Fb \Fb \Pb \Pb,\bbS}+ S_{\Fb \Pb \Fb \Pb,\bbS}+ S_{\Fb \Pb \Pb \Pb,\bbS} \ ,
    \end{split}
\end{equation}
where the subscripts denote the number of different boundary sources. Again (similar to the three-point influence phase), the terms with only future or past sources, vanish, i.e., $S_{ \Fb \Fb \Fb \Fb}=0=S_{ \Pb \Pb \Pb \Pb}$. This means the SK collapse and the KMS condition also hold at the four-point tree level.

Here, we will not give all the terms of the four-point Yukawa influence phase, but only explicitly write down the $S_{\Fb \Fb \Fb \Pb,\bbG}$ term. The interested reader is referred to Appendix (\ref{sec:AppendixFourPointcolumnVector}) for the full set of terms.
{\small
\begin{equation}
    \begin{split}
        &S_{\Fb \Fb \Fb \Pb,\bbG} =-  \int_{k_{1,2,3,4}} \Big\{ \langle \Sbin \Sin \bbGA \Sbin  \Sout \rangle \cdot   \fF^{{}_{(1)}}(\bar{k}_1)  \otimes \fF^{{}_{(1)}}(\bar{k}_2) \otimes \fF^{{}_{(1)}}(\bar{k}_3)  \otimes \left[ \fP^{{}_{(2)}}(\bar{k}_4)-\fP^{{}_{(1)}}(\bar{k}_4)\right]\\
        &\hspace{3cm}+\langle  \Sbin \Sin \bbGA \Sbout  \Sin  \rangle \cdot  \fF^{{}_{(1)}}(\bar{k}_1)  \otimes \fF^{{}_{(1)}}(\bar{k}_2) \otimes \left[ \fP^{{}_{(2)}}(\bar{k}_3)-\fP^{{}_{(1)}}(\bar{k}_3)\right]  \otimes \fF^{{}_{(1)}}(\bar{k}_4) \Big\} \ ,
    \end{split}
\end{equation}
}
where we have defined
{\small
\begin{equation}
    \begin{split}
        \langle \overline{S}^{x_1} S^{x_2}& \bbG_{x} \overline{S}^{x_3}  S^{x_4} \rangle \equiv \int_{\text{ext}_1} \int_{\text{ext}_2} \left[\overline{S}^{x_1}(\zeta_2, k_1)\right]^{\alpha}_{\ \bulkalpha} \left[S^{x_2}(\zeta_2, k_2)\right]^{\bulkalpha}_{\ \beta} \bbG_{x}(\zeta_2|\z_1, k_3+k_4) \\
        & \hspace{3cm} \times  \left[\overline{S}^{x_3}(\zeta_1, k_3)\right]^{\gamma}_{\ \bulkbeta}\left[S^{x_4}(\zeta_1, k_4)\right]^{\bulkbeta}_{\ \delta}  \left[ \boldsymbol{\psibar}(k_1)  \right]_{\alpha} \otimes \left[\boldsymbol{\psi}(k_2)\right]^{\beta} \otimes \left[ \boldsymbol{\psibar}(k_3)  \right]_{\gamma} \otimes \left[\boldsymbol{\psi}(k_4)\right]^{\delta}\ .
    \end{split}
    \label{eq:DefBraketNotationSpectralFunc}
\end{equation}
}
where $x_1, x_2, x_3 \in \{\text{in}, \text{out}\}$ and $x \in \{\text{ret}, \text{adv}\}$ . Notice once again, as in the case of the three-point functions, that all the statistical factors are naturally incorporated into the column vectors.

\section{Discussion}\label{sec:Discussion}

In this work, we have constructed an open effective field theory for a finite-temperature holographic system comprising interacting fermions. This was achieved by employing the gravitational dual of the Schwinger-Keldysh formalism. While this formalism had been previously used for free fermions \cite{Loganayagam:2020eue, Loganayagam:2020iol}, our contribution lies in its application to interacting fermions for the first time.

In the bulk, our framework derives an exterior field theory in a black hole background. The setup in the bulk involves a massless scalar and a massless Dirac field with a Yukawa interaction in the AdS black hole background. Additionally, we have developed Witten diagrammatics for this exterior field theory which allows us to write down the Schwinger-Keldysh correlators of the boundary theory. We have provided explicit expressions for these correlators up to four-point functions, at tree level in the bulk, using Witten diagrams. However, the Witten diagrammatics we provide can be used to compute the tree-level influence phase to arbitrary orders in the Yukawa coupling constant.

Another contribution of our work is the derivation of the column vector representation for fermions. This representation facilitates the concise expression of correlators wherein the Kubo-Martin-Schwinger (KMS) conditions and the Schwinger-Keldysh collapse conditions become manifest.

This work has many clear future directions. One of them is to understand quantum electrodynamics (QED) in the gravitational Schwinger-Keldysh geometry. Now that we understand how to deal with fermions interacting with scalars, this will be the next natural step. The gauge field of QED can be decomposed into (designer) scalars as in \cite{Ghosh:2020lel}, and the QED vertex will reduce to interactions between these scalars and the fermions. Thus, it will be interesting to see how our work generalises to the case of designer scalars, and thus gauge fields. We intend to explore this question in the future.

Physically, in the bulk, this corresponds to the question of photons and matter scattering against each other in a black hole background. More precisely, how does the Hawking radiation of the black hole, comprising photons and matter, affect the scattering of particles thrown in from the boundary? This question becomes particularly interesting in the case of charged black holes \cite{Loganayagam:2020iol}, since now the black hole couples to fields not only gravitationally but also electromagnetically.

Another future direction is to study the divergences present in the many bulk integrals in this work. An interesting question would be to ask whether the prescription of source-renormalisation, first detailed in \cite{Jana:2020vyx}, becomes important in the present case. This will be an independent check of the source-renormalisation prescription. 

Throughout this work, we have restricted ourselves to computing the bulk on-shell action, which is equivalent to the tree-level influence phase. It would be interesting to explore the loop contributions in the bulk. These loop corrections correspond to the large-$N$ corrections in the boundary field theory. We have some preliminary results in this direction which we hope to present soon.

It is also interesting to ask how much of the analysis presented here generalises to the case of de Sitter spacetime. In the spirit of \cite{Sleight:2021plv, Schaub:2023scu, Bhattacharya:2023twz}, we ask what our analysis teaches us about cosmological correlators in the Schwinger-Keldysh formalism.

\section*{Acknowledgements}
We express our sincere gratitude to Subhro Bhattacharjee, Chandramouli Chowdhury, R. Loganayagam, Gautam Mandal, Shiraz Minwalla, Suvrat Raju, Mukund Rangamani, Tamoghna Ray, Omkar Shetye, Akhil Sivakumar, and the ICTS string group for valuable discussions. Furthermore, SKS would like to thank the TIFR string group for allowing them to present this work there. SKS would also like to thank the organizers of the \textit{Future Perspectives on QFT and Strings} conference held at IISER, Pune, where this work was presented. We acknowledge the support of the Department of Atomic Energy, Government of India, under project no. RTI4001, and the unwavering and generous support provided by the people of India towards research in the fundamental sciences.

\appendix

\section{Conjugate Dirac equation and its propagators}\label{app:ConjDirac}

In this appendix, we will discuss the conjugate Dirac equation and its solutions. Varying the Dirac action (\ref{eq:Diracaction}) with respect to the field $\Psi$ gives us the conjugate Dirac equation:
\begin{equation}\label{eq:Diracbar}
    \left(r^{d/2} \Psibar(\zeta,k) \right) \left\{ \overleftarrow{\partial}_\zeta \frac{\bbT(\zeta)}{\sqrt{f}} + \mathbb{\Gamma} \left(\beta k^0 + \frac{1}{2} \partial_\zeta \ln f \right) -\frac{\beta}{2} \Gamma^{(i)} k_i \right\} = 0 \ ,
\end{equation}
where we have already passed to the Fourier domain (in the boundary coordinates) by writing the conjugate field as
\begin{equation}
    \Psibar(\zeta,x) \equiv \int_{k}  e^{ikx} \  \Psibar (\zeta, k)\ .
\end{equation}
Here the leftward pointing arrow on the radial derivative denotes that the derivative acts on the object that is to its left. Note that this conjugate equation (and its solutions) can also be derived by performing the conjugation operation on the Dirac equation, i.e., by using $\Psibar \equiv \Psi^\dag \Gamma^{(v)}$. But here we will treat it as an independent equation and solve it accordingly. 

The boundary conditions for the conjugate field are taken to be the conjugated version of those given for the Dirac field Eq.~(\ref{eq:BCDiracSol}). Given the right and the left boundary conjugate sources, $\psibar_{\sR}$ and $\psibar_{\sL}$ respectively, the boundary conditions on the bulk conjugate field $\Psibar$ take the form
\begin{equation}\label{eq:BCDiracbar}
    \lim_{\zeta \to 0} r^{\frac{d}{2}} \Psibar(\zeta,k) \projm = \psibar_{\sL}(k) \Sbo \projm \  , \qquad 
    \lim_{\zeta \to 1} r^{\frac{d}{2}}  \Psibar(\zeta,k) \projm = \psibar_{\sR}(k) \Sbo \projm \ ,
\end{equation}
where we have defined 
\begin{equation}
    \psibar(k) \equiv \psi^\dag(k) \gamma^{(v)} \, \quad \text{and} \quad \Sbo \equiv \left[ \gamma^{(v)} \right]^\dag \left[S_0\right]^\dag \  \Gamma^{(v)} \ .
\end{equation}
Here $\gamma^{(v)}$ is a boundary gamma matrix.

We can now proceed with the usual procedure: find ingoing and outgoing propagators and then build the full solution from them. The ingoing boundary-to-bulk conjugate propagator $\Sbin$ satisfies the boundary conditions
\begin{equation}\label{eq:BCSbin}
    \lim_{\zeta \to 0}r^{\frac{d}{2}}  \Sbin(\zeta,k) \projm =   \Sbo \projm=\lim_{\zeta \to 1}r^{\frac{d}{2}}  \Sbin(\zeta,k)  \projm  \ .
\end{equation}
The time-reversal involution of the grSK geometry then produces the outgoing boundary-to-bulk conjugate propagator $\Sbout$, given by\footnote{The negative sign is convenient at the level of the Feynman rules for the Witten diagrams.}
\begin{equation}
    \Sbout(\z,k) = - e^{-\beta k^0 \z} \  \Sbin(\z,-k) \mathbb{T}(\z) \ .
\end{equation}
The boundary limits of the outgoing conjugate propagator $\Sbout(\z,k)$ defined above, can also be obtained using Eq.~\eqref{eq:BCSbin}. Now, using the ingoing and the outgoing boundary-to-bulk propagators constructed above, the complete solution of the conjugate Dirac equation can be written as
\begin{equation}\label{eq:ConjSol}
    \Psibar(\zeta,k) =  \ -\psibar_{\Fb}(k) \Sbin (\zeta,k) - \psibar_{\Pb}(k) \Sbout(\zeta,k) e^{\beta k^0}  \ .
\end{equation}
Here we have defined the conjugate boundary sources in the past-future basis as
\begin{equation}
    \psibar_{\Fb}(k) \equiv n^{\fd}_{k^0} \left[ \psibar_{\sR}(k) -\psibar_{\sL}(k)\right] - \psibar_{\sR}(k) \ , \quad \psibar_{\Pb}(k) \equiv n^{\fd}_{k^0} \left[\psibar_{\sR}(k) - \psibar_{\sL}(k) \right]\ .
\end{equation}

In the right-left basis, the solution takes the form
\begin{equation}
    \Psibar(\zeta,k) = \psibar_{\sR}(k) \overline{s}_{\sR}(\zeta,k) - \psibar_{\sL}(k) \overline{s}_{\sL}(\zeta,k)  \ ,
\end{equation}
where
\begin{equation}\label{eq:sRbarsLbarSinbar}
    \begin{split}
   \overline{s}_{\sR} (\zeta, k) &\equiv (1-n^{\fd}_{k^0}) \left\{\Sbin (\zeta,k) - \Sbout(\zeta,k) \right\} \ ,\\
    \overline{s}_{\sL} (\zeta, k) &\equiv (1-n^{\fd}_{k^0})\left\{-e^{-\beta k^0} \Sbin(\zeta,k) -  \Sbout(\zeta,k) \right\}\ .\\
    \end{split}
\end{equation}

\section{Bulk-to-bulk propagators in the grSK geometry}
In this section, we will discuss the derivation of bulk-to-bulk propagators in the grSK geometry. We will begin with a quick review of scalar bulk-to-bulk propagators and how they are obtained. Here we will mostly follow the techniques used in \cite{Arnold:2011hp, Faulkner:2013bna} and also implemented for the grSK geometry in \cite{Loganayagam:2024mnj}.

Following this discussion on scalars, we will come to a detailed derivation of the spinor bulk-to-bulk propagator in the grSK geometry. 

\subsection*{Scalar bulk-to-bulk propagator}

The scalar bulk-to-bulk propagator $\bbG(\z|\z_0,k)$ normalised to unity at the left and the right boundary (henceforth binormalisable) is given by
\begin{equation}
    \bbG(\z|\z_0,k) = \red{-}\frac{e^{\beta k^0 \z_0} }{1+n^{\be}_{k^0}}\frac{g_{\sL}(\z_>,k) g_{\sR}(\z_<,k) }{\Kin(k)-\Kin(-k)} \ .
\end{equation}
Here the symbols $\zeta_<$ and $\zeta_>$ are defined by
\begin{equation}
    \zeta_< \equiv \Bigl\{ 
        \begin{array}{cc}
            \zeta & \text{if $\zeta$ comes before $\zeta_0$ on the grSK contour}\\
            \zeta_0 & \text{if $\zeta_0$ comes before $\zeta$ on the grSK contour} \ ,
        \end{array}
    \label{zetalesserDef}
\end{equation}
and
\begin{equation}
    \zeta_> \equiv \Bigl\{ 
        \begin{array}{cc}
            \zeta & \text{if $\zeta$ comes after $\zeta_0$ on the grSK contour}\\
            \zeta_0 & \text{if $\zeta_0$ comes after $\zeta$ on the grSK contour} \ .
        \end{array} 
    \label{zetagreaterDef}
\end{equation}
Since the bulk-to-bulk propagator has to solve the free Klein-Gordon equation away from the sources, it is clear that it has to be proportional to the boundary-to-bulk propagators. Left and right-normalisability then fixes this combination of the boundary-to-bulk propagators to be $g_{\sL}(\z_>,k) g_{\sR}(\z_<,k)$. The other factors in the above expression must be fixed using the discontinuity condition coming from the delta function source in the Klein-Gordon equation. Here, $\Kin$ is the retarded boundary two-point correlator, obtained by taking the boundary limit of the ingoing boundary-to-bulk propagator as shown in \cite{Ghosh:2020lel}.

Apart from the binormalisable bulk-to-bulk propagator, we also require the retarded and the advanced bulk-to-bulk propagators, given by
\begin{equation}\label{eq:DefScalarBlkBlkRetAdv}
    \begin{split}
        \bbGR(\z|\z_0,k) &=  \frac{e^{\beta k^0 \z_0} }{1+n^{\be}_{k^0}}\frac{\Gin(\z_>,k) g_{\sR}(\z_<,k) }{\Kin(k)-\Kin(-k)} \ , \\  
        \bbGA(\z|\z_0,k) &= \frac{e^{\beta k^0 \z_0} }{1+n^{\be}_{k^0}}\frac{\Gout(\z_>,k) g_{\sR}(\z_<,k) }{\Kin(k)-\Kin(-k)} \ . 
    \end{split}
\end{equation}
These are both propagators that are normalisable at the left boundary. Furthermore, $\bbGR$ is analytic in the upper half-plane (UHP) of frequency of $k^0$, and $\bbGA$ is analytic in the lower half-plane (LHP) frequency of $k^0$. We will now present an argument to show these analyticity properties. 

We begin by detailing the analyticity properties of the ingoing and the outgoing boundary-to-bulk propagators. The boundary-to-bulk propagators, $\Gin$ and $\Gout$, following \cite{Loganayagam:2022zmq}, can be written as
\begin{equation}
    \Gin(\z,k) = \Kin(k) \tGin(\z,k)  \ , \qquad \Gout(\z,k) = \Kin(-k) \tGout(\z,k) \ ,
\end{equation}
where $\tGin$ and  $\tGout$ are generically regular\footnote{This is easy to see in $d=2$, from the explicit expressions of these propagators provided in \cite{Jana:2020vyx}.} functions of $k^0$. Here we have used the intuition that the ingoing propagator should have poles only at quasinormal frequencies \cite{Birmingham:2001pj}. It also highlights the fact that $\Gin$ is analytic in the UHP of $k^0$.

Coming to the retarded and the advanced bulk-to-bulk propagators in Eq.~\eqref{eq:DefScalarBlkBlkRetAdv}, as far as analyticity in the $k^0$ plane is concerned, we can replace $\left[ \Kin(k)-\Kin(-k) \right]$ by $\left[ \Kin(k) \Kin(-k) \right]$. This leads to
\begin{equation}
    \bbGR (k)= \Kin(k) \tbbGR (k) \ , \qquad \bbGA (k)= \Kin(-k) \tbbGA(k) \ ,
\end{equation}
where $\tbbGR$ and  $\tbbGA$ are again generically regular functions of $k^0$. Therefore $\bbGR$ and $\bbGA$ are analytic in the upper and the lower halves of the $k^0$ plane respectively.

\subsection*{Spinor bulk-to-bulk propagators}\label{sec:DerivationSpinorialBlkBlk}

We now come to the spinor bulk-to-bulk propagators. The computation of the spinor bulk-to-bulk propagator will require the evaluation of a quantity $\bbW(\zeta,k)$ which satisfies $\partial_\zeta \bbW(\zeta,k) =0$, i.e., the analogue of the Wronskian of a second-order ODE. We will, therefore, spend some time in computing this quantity explicitly and then use it to compute the spinor bulk-to-bulk propagator.

\subsubsection*{Derivation of the conserved quantity}\label{sec:ConsQuant}
Consider $\Psibar_2$ and $\Psi_1$, which solve the conjugate Dirac equation and the Dirac equation, given by
\begin{equation}
    \left(r^{d/2} \Psibar_2(\zeta,k) \right) \left\{ \overleftarrow{\partial}_\zeta \frac{\bbT(\zeta)}{\sqrt{f}} + \mathbb{\Gamma} \left(\beta k^0 + \frac{1}{2} \partial_\zeta \ln f \right) -\frac{\beta}{2} \Gamma^{(i)} k_i \right\} = 0 \ ,
    \label{eq:StepsWronskianEval1}
\end{equation}
and
\begin{equation}
    \left\{ \frac{\bbT(\zeta)}{\sqrt{f}}\partial _\zeta + \mathbb{\Gamma} \left(\beta k^0 + \frac{1}{2} \partial_\zeta \ln f \right) - \frac{\beta}{2} \Gamma^{(i)} k_i  \right\}\left(e^{-\beta k^0 \zeta} \bbT(\zeta) r^{d/2} \Psi_1 (\zeta,-k) \right)  = 0 \ ,
    \label{eq:StepsWronskianEval2}
\end{equation}
respectively.
Note that if $\Psi_1(\zeta,k)$ is a solution of the Dirac equation, then so is its time reversal $e^{-\beta k^0 \zeta} \bbT(\zeta) r^{d/2} \Psi_1 (\zeta,-k) $, which is what we have used in the second equation above. Multiplying Eq.~\eqref{eq:StepsWronskianEval1} from the right by $\sqrt{f} e^{-\beta k^0 \zeta} r^{d/2} \Psi_1 (\zeta,-k)$, Eq.~\eqref{eq:StepsWronskianEval2} from the left by $\sqrt{f}  r^{d/2} \Psibar_2(\zeta,k) \bbT(\zeta)$ and adding, we obtain\footnote{Here, we have used the bulk gamma matrix identities
\begin{equation}
    \Gamma^{(i)} + \bbT(\zeta) \Gamma^{(i)} \bbT(\zeta) = 0 \ , \quad \text{and} \quad \sqrt{f} \left[\mathbb{\Gamma} + \bbT(\zeta) \mathbb{\Gamma} \bbT(\zeta) \right] = \bbT(\zeta) \ .
\end{equation}}
\begin{equation}
    \begin{split}
        &\partial_{\zeta} \left\{r^{d/2} \Psibar_2(\zeta,k) \bbT(\zeta) e^{-\beta k^0 \zeta} r^{d/2} \Psi_1 (\zeta,-k) \right\}\\
        &\hspace{2cm}+ \left(\beta k^0 + \frac{1}{2} \partial_\zeta \ln f \right) \left\{r^{d/2} \Psibar_2(\zeta,k) \bbT(\zeta) e^{-\beta k^0 \zeta} r^{d/2} \Psi_1 (\zeta,-k) \right\} = 0  \ ,
    \end{split}
\end{equation}
which integrates to
\begin{equation}
    \partial_\zeta \bbW(\zeta,k) =0 \ , \quad \text{where} \quad \bbW(\zeta,k) \equiv  \sqrt{f} r^{d/2} \Psibar_2(\zeta,k) \bbT(\zeta) r^{d/2} \Psi_1 (\zeta,-k)  \ .
\end{equation}
Note that this conserved quantity contains the boundary sources. To construct a bulk-to-bulk propagator we would require a quantity independent of boundary sources. So, we now define a square matrix $\frakW(\z,k)$ in the boundary spinor space where we strip off the boundary sources from this conserved quantity $\bbW(\z,k)$, given by
\begin{equation}\label{eq:DefbbW}
    \frakW[\bar{s}_2(\zeta,k), s_1(\zeta,k)] \equiv  \sqrt{f} r^{d/2} \bar{s}_2(\zeta,-k) \bbT(\zeta) r^{d/2} s_1 (\zeta,k) \ .
\end{equation}
Here, $\bar{s}_2(\zeta,k)$ and $s_1(\zeta,k)$ are the conjugate boundary-to-bulk and boundary-to-bulk propagators respectively.

This boundary spinor-space matrix in the basis of ingoing and outgoing solutions takes the form
\begin{equation}
    \begin{split}
        \lim_{\zeta \to 0,1} \frakW[\Sbin(\zeta,k), \Sin(\zeta,k)] &= \lim_{\zeta \to 0,1} \sqrt{f} r^{d/2} \Sbin(\zeta,-k) \bbT(\zeta)r^{d/2} \Sin(\zeta,k)\\
        &= \lim_{\zeta \to 0,1} r^{d/2} \Sbin(\zeta,-k) (\projp-\projm) r^{d/2} \Sin(\zeta,k)\\
        &= -\left\{\Kret(k)+\Kret(-k) \right\} \ ,
    \end{split}
\end{equation}
where we have used Eq.~\eqref{eq:BCSin}, Eq.~\eqref{eq:BCSbin} and have also defined the boundary retarded two-point correlator $\Kret(k)$ as
\begin{equation}\label{eq:Kret}
    \Kret(k) \equiv \lim_{\z \to 0,1}\Sbo \projm r^{d/2} \Sin(\zeta,k) = -\lim_{\zeta \to 0,1} r^{d/2} \Sbin(\z,k) \projp S_0 \ .
\end{equation}
By similar computations, we find the other combinations:
\begin{equation}
    \begin{split}
        &\frakW[\Sbin(\zeta,k), \Sin(\zeta,k)] =  -\left\{\Kret(k) +\Kret(-k) \right\} = -\frakW[\Sbout(\zeta,k), \Sout(\zeta,k)] \ ,\\
        &\frakW[\Sbout(\zeta,k), \Sin(\zeta,k)] = 0 = \frakW[\Sbin(\zeta,k), \Sout(\zeta,k)] \ .
    \end{split}
\end{equation}
Similar expressions for the conserved quantity corresponding to the right and the left boundary-to-bulk propagators can be derived from the above equations using Eq.~\eqref{eq:sRsLSin} and Eq.~\eqref{eq:sRbarsLbarSinbar}. Explicitly, we have
\begin{equation}\label{eq:WronsRLbasis}
    \begin{split}
        &\frakW[\bar{s}_{\sR}(\zeta,k), s_{\sR}(\zeta,k)] = 0 = \frakW[\bar{s}_{\sL}(\zeta,k), s_{\sL}(\zeta,k)] \ ,\\
        &\frakW[\bar{s}_{\sR}(\zeta,k), s_{\sL}(\zeta,k)] = n^{\fd}_{k^0} \left\{\Kret(k)+\Kret(-k) \right\} = e^{-\beta k^0} \frakW[\bar{s}_{\sL}(\zeta,k), s_{\sR}(\zeta,k)] \ .
    \end{split}
\end{equation}

\subsubsection*{Determining the spinor bulk-to-bulk propagator}
Let us now take a look at the differential equation satisfied by the spinorial bulk-to-bulk propagator (Eq.~\eqref{eq:bbSODE}). Away from the point $\zeta = \zeta_0$, it is simply the free Dirac equation and therefore in these regions, the bulk-to-bulk propagator has to be proportional to the boundary-to-bulk propagators. Keeping in mind the boundary conditions in Eq.~\eqref{eq:BCbbS}, we write an ansatz for the bulk-to-bulk propagator as
\begin{equation}
    \begin{split}
        \bbS(\zeta|\zeta_0,k) &=  \left[ {s}_{\sR} (\zeta,k) M_<(\zeta_0,k)  \bar{s}_{\sL}(\zeta_0,-k) \right]  \thetaSK(\zeta<\zeta_0)\\
        &\hspace{0.05cm} +  \left[ {s}_{\sL} (\zeta,k) M_>(\zeta_0,k)  \bar{s}_{\sR}(\zeta_0,-k) \right]  \thetaSK(\zeta>\zeta_0)  \ .
    \end{split}
    \label{eq:BlkBlkAnsatz}
\end{equation}
Here, the symbol $\thetaSK$ denotes a Heaviside theta function on the grSK contour defined as
\begin{equation}
    \begin{split}
        &\thetaSK(\zeta <\zeta_0) \equiv\Bigg\{
            \begin{array}{cc}
                \zeta & \text{if $\zeta$ comes before $\zeta_0$ on the grSK contour}\\
                \zeta_0 & \text{if $\zeta_0$ comes before $\zeta$ on the grSK contour} 
            \end{array}\ ,\\
        &\thetaSK(\zeta >\zeta_0) \equiv 1- \thetaSK(\zeta <\zeta_0) \ .
    \end{split}
\end{equation}
The symbols $M_<(\zeta_0,k)$ and $M_>(\zeta_0,k)$ denote square matrices in the boundary spinor space which we will soon determine using the jump conditions at $\zeta = \zeta_0$. It is easy to see that the above ansatz is a square matrix in the bulk spinor space and also that it satisfies the binormalisable boundary conditions given in Eq.~\eqref{eq:BCbbS}.

The jump condition is obtained by integrating the delta-sourced Dirac equation in Eq.\eqref{eq:bbSODE} around the point $\zeta= \zeta_0$. Explicitly, we have
\begin{equation}
    \begin{split}
        \frac{i \beta r}{2} r^{d/2} \Gamma^{\A} D_{\A} \bbS(\zeta|\zeta_0,k) &= \left\{\frac{\bbT(\zeta)}{\sqrt{f}} \partial_\zeta + \mathbb{\Gamma} \left(\beta k^0 + \frac{1}{2} \partial_\zeta \ln f \right) - \frac{\beta}{2} \Gamma^{(i)} k_i  \right\}\left(r^{d/2} \bbS(\zeta|\zeta_0,k) \right)\\
        & = \red{-}\mathbb{1} \frac{\delta(\zeta-\zeta_0)}{r^{d/2} f} \ ,
    \end{split}
\end{equation}
where we have used the fact that $\sqrt{-g} = \frac{i \beta }{2} r^{d+1} f$.
Integrating the above ODE from $\zeta = \zeta_0^-$ to $\zeta= \zeta_0^+$, we arrive at the jump condition
\begin{equation}
    \frac{\bbT(\zeta)}{\sqrt{f}} r^{d/2} \bbS (\zeta|\zeta_0,k) \Bigg|_{\zeta = \zeta_0^-}^{\zeta = \zeta_0^+} = \red{-}\frac{\mathbb{1}}{r_0^{d/2} f(\zeta_0)} \ .
\end{equation}
Using the ansatz in Eq.~\eqref{eq:BlkBlkAnsatz}, the above jump condition takes the form
\begin{equation}
    \sqrt{f} r^d \bbT(\zeta)\left\{ {s}_{\sL}(\zeta,k) \  M_> (\zeta,k) \ \bar{s}_{\sR}(\zeta,-k) - {s}_{\sR}(\zeta,k) \ M_< (\zeta,k)  \ \bar{s}_{\sL}(\zeta,-k)  \right\} = \red{-}\mathbb{1} \ .
\end{equation}
Notice now that pre-multiplying and post-multiplying the above equation with appropriate boundary-to-bulk propagators, we obtain simple equations involving the conserved quantity that we computed in Eq.~\eqref{eq:DefbbW} and Eq.~\eqref{eq:WronsRLbasis}. Pre-multiplying by $r^{d/2}\bar{s}_{\sL}(\zeta,-k)$ and post-multiplying by $\sqrt{f}\bbT r^{d/2} s_{\sR}(\zeta,k)$, we obtain $M_<(\zeta,k)$. Similarly, pre-multiplying by $r^{d/2}\bar{s}_{\sR}(\zeta,-k)$ and post-multiplying by $\sqrt{f}\bbT r^{d/2} s_{\sL}(\zeta,k)$, we obtain $M_>(\zeta,k)$. The explicit expressions are as follows.
\begin{equation}
    M_<(\zeta,k) = \frac{1}{(1-n^{\fd}_{k^0})\left\{\Kret(k)+\Kret(-k) \right\}} \ ,\quad M_>(\zeta,k) = \red{-}\frac{1}{n^{\fd}_{k^0}\left\{\Kret(k)+\Kret(-k) \right\}} \ .
\end{equation}

Substituting these results in the ansatz in Eq.~\eqref{eq:BlkBlkAnsatz}, we obtain
\begin{equation}
    \begin{split}
        \bbS(\zeta|\zeta_0,k) &=  \left[ {s}_{\sR} (\zeta,k) \frac{1}{(1-n^{\fd}_{k^0})\left\{\Kret(k)+\Kret(-k) \right\}} \bar{s}_{\sL}(\zeta_0,-k) \right]  \thetaSK(\zeta<\zeta_0)\\
        &\quad \red{-}  \left[ {s}_{\sL} (\zeta,k) \frac{1}{n^{\fd}_{k^0} \left\{\Kret(k)+\Kret(-k) \right\}}  \bar{s}_{\sR}(\zeta_0,-k) \right]  \thetaSK(\zeta>\zeta_0)  \ .
    \end{split}
\end{equation}
We take a moment to also note that the $\Pmin$ projected boundary limits of the above propagator take the form
\begin{equation}
    \lim_{\z \to 0}r^{\frac{d}{2}}\Sbo \Pmin \bbS(\z|\z_0,k) = \overline{s}_{\sL}(\z_0,k) \ , \qquad  \lim_{\z \to 1}r^{\frac{d}{2}}\Sbo \Pmin \bbS(\z|\z_0,k) = \overline{s}_{\sR}(\z_0,k) \ .
    \label{eq:BoundaryLimitsSpinorBlkBlk}
\end{equation}

\subsection*{Retarded and advanced spinor bulk-to-bulk propagators}
Now that we have the binormalisable bulk-to-bulk propagator, we can also construct bulk-to-bulk propagators that have specific causal properties, i.e., the retarded and the advanced bulk-to-bulk propagators. These propagators are defined to be normalisable at the left boundary, and analytic in the upper-half and the lower-half of the frequency plane respectively. Here, we will not provide a derivation of these propagators, as it is similar to the computation of the binormalisable bulk-to-bulk propagator given above. We just quote the results:
\begin{equation}
    \begin{split}
        \bbSR(\zeta|\zeta_0,k) &=  \red{-}\left[ {s}_{\sR} (\zeta,k) \frac{1}{(1-n^{\fd}_{k^0})\left\{\Kret(k)+\Kret(-k) \right\}} \Sbout(\zeta_0,-k) \right]  \thetaSK(\zeta<\zeta_0)\\
        &\quad + \left[ \Sin(\zeta,k) \frac{1}{n^{\fd}_{k^0} \left\{\Kret(k)+\Kret(-k) \right\}}  \bar{s}_{\sR}(\zeta_0,-k) \right]  \thetaSK(\zeta>\zeta_0)  \ ,
    \end{split}
    \label{eq:DefRetardedSpinorBlkBlk}
\end{equation}
and
\begin{equation}
    \begin{split}
        \bbSA(\zeta|\zeta_0,k) &=  \red{-}\left[ {s}_{\sR} (\zeta,k) \frac{1}{(1-n^{\fd}_{k^0})\left\{\Kret(k)+\Kret(-k) \right\}} \Sbin(\zeta_0,-k) \right]  \thetaSK(\zeta<\zeta_0)\\
        &\quad + \left[ \Sout(\zeta,k) \frac{1}{n^{\fd}_{k^0} \left\{\Kret(k)+\Kret(-k) \right\}}  \bar{s}_{\sR}(\zeta_0,-k) \right]  \thetaSK(\zeta>\zeta_0)  \ .
    \end{split}
    \label{eq:DefAdvancedSpinorBlkBlk}
\end{equation}

Now, we show how these propagators satisfy all the required properties. First of all, we note that the right boundary-to-bulk propagator $s_{\sR}(\zeta,k)$ satisfies
\begin{equation}
    \lim_{\z \rightarrow 0} r^{\frac{d}{2}} \Pplus {s}_{\sR}(\z,k) = 0  \ .
\end{equation}
This implies that the retarded and the advanced bulk-to-bulk propagators satisfy
\begin{equation}
    \lim_{\z \rightarrow 0} r^{\frac{d}{2}} \Pplus \bbSR(\zeta|\zeta_0,k) = 0  \  , \qquad \lim_{\z \rightarrow 0} r^{\frac{d}{2}} \Pplus \bbSA(\zeta|\zeta_0,k) = 0 \ .
\end{equation}

Next, we have to show that the above propagators have the desired analyticity properties on the complex frequency plane. To do this, we will employ a similar set of arguments as the one used in the case of the scalar bulk-to-bulk propagator. To begin with, we use the intuition that the ingoing boundary-to-bulk propagator $\Sin(\zeta,k)$ has the same pole structure, i.e., the quasinormal poles, on the complex frequency plane as the boundary retarded correlator $\mathfrak{K}_{\rm ret}(k)$. We can then use the same trick we used for the case of the scalar retarded and advanced propagators to conclude that $\bbSR(\zeta|\zeta_0,k)$, has the same poles as $\mathfrak{K}_{\rm ret}(k)$ and is analytic in the upper half of the frequency plane. Similarly, $\bbSA(\zeta|\zeta_0,k)$ is analytic in the lower half of the frequency plane.

Here, we point out that the binormalisable spinor bulk-to-bulk propagator can be written as a linear combination of the retarded and the advanced bulk-to-bulk propagators as
\begin{equation}
    \bbS(\z|\z_0,k) =  \  n^{\fd}_{k^0} \bbSR(\z|\z_0,k)+ (1- n^{\fd}_{k^0})\bbSA(\z|\z_0,k) \ .
\end{equation}
This relation can be easily obtained by replacing the corresponding propagators, and $n^{\be}_{k^0}$ with $-n^{\fd}_{k^0}$, from the corresponding relation for scalar bulk-to-bulk propagators \cite{Loganayagam:2024mnj}:
\begin{equation}
    \bbG(\z|\z_0,k) =  \  -n^{\be}_{k^0} \bbGR(\z|\z_0,k)+ (1+ n^{\be}_{k^0})\bbGA(\z|\z_0,k) \ .
\end{equation}

\section{Monodromy integrals around the grSK contour}\label{Sec:AppendixMonodromyIntegrals}
The computation of the four-point exchange diagrams that we have detailed in the main text requires performing integrals of the form
\begin{equation}
    \mathscr{I}_{\bbG} \equiv \oint_{\zeta_1} \oint_{\zeta_2} \ [f(\zeta_1)]^{\frac{\alpha_1}{2}} [f(\zeta_2)]^{\frac{\alpha_2}{2}}  e^{-\beta k^0_1  \zeta_1}  e^{-\beta k^0_2  \zeta_2}  \bbG(\zeta_2|\zeta_1, p) \mathscr{F}(\zeta_1, \zeta_2) \ ,
\end{equation}
and
\begin{equation}
    \mathscr{I}_{\bbS} \equiv \oint_{\zeta_1} \oint_{\zeta_2} \ [f(\zeta_1)]^{\frac{\alpha_1}{2}} [f(\zeta_2)]^{\frac{\alpha_2}{2}} e^{-\beta k^0_1  \zeta_1}  e^{-\beta k^0_2  \zeta_2}  \bar{\mathfrak{a}}_2(\zeta_2)\bbS(\zeta_{2}|\zeta_{1}, p) \mathfrak{a}_1(\zeta_1)  \ .
\end{equation}
Here $\alpha_1$ and $\alpha_2$ are integers, $\mathscr{F}(\zeta_1,\zeta_2)$ is a scalar analytic function on the complex $r_1$ and $r_2$ planes, and $\mathfrak{a}_1(\zeta)$ and $\bar{\mathfrak{a}}_2(\zeta)$ are analytic functions on the $r$ plane that has the spinor structure of a spinor and a conjugate spinor boundary-to-bulk propagator respectively.

The double-discontinuity integrals above reduce to integrals purely on one copy of the black hole exterior. That is, the monodromy around the grSK branch cut can be explicitly computed and we can write the answer for the contour integral as an integral over the real $r$ line. In this appendix, we will show this in detail for the second integral above, i.e., the one involving the spinor bulk-to-bulk propagator, $\mathscr{I}_{\bbS}$.

On the other hand, the double discontinuity integral of the scalar bulk-to-bulk propagator has been computed in \cite{Loganayagam:2024mnj}. Here the authors were concerned only with scalar interactions and thus do not include the factors of $\sqrt{f}$ we have above. Nevertheless,  including the new factors of $\sqrt{f}$ is relatively straightforward and we just present the result here.
\begin{equation}
    \begin{split}
        \mathscr{I}_{\bbG} = \int_{\rm ext_1} \int_{\rm ext_2} [f(\zeta_1)]^{\frac{\alpha_1}{2}} [f(\zeta_2)]^{\frac{\alpha_2}{2}} e^{-\beta k^0_1 \zeta_1} e^{-\beta k^0_2 \zeta_2} \bbG_{\rm DD}(\zeta_2|\zeta_1,p) \mathscr{F}(\zeta_1,\zeta_2) \ ,
    \end{split}
    \label{eq:DoubleDiscScalar}
\end{equation}
where $\int_{\rm ext}$ denotes the integral over the exterior of the left copy of the black bole, defined in Eq.~\eqref{eq:DefExteriorIntegral}. The symbol $\bbG_{\rm DD}(\zeta_2|\zeta_1,p)$ denotes the double-discontinuity given by 
\begin{equation}
    \begin{split}
        \bbG_{\rm DD}(\zeta_2|\zeta_1,p) &\equiv \frac{-n^{\be}_{p^0}}{\left(1+\frac{1}{(-)^{\alpha_1} e^{\beta (k_1^0 - p^0)}-1}\right)\left(1+\frac{1}{(-)^{\alpha_2} e^{\beta k_2^0}-1}\right)} \bbGR(\zeta_2|\zeta_1,p)\\
        & +\frac{1+n^{\be}_{p^0}}{\left(1+\frac{1}{(-)^{\alpha_1} e^{\beta k_1^0}-1}\right)\left(1+\frac{1}{(-)^{\alpha_2} e^{\beta (k_2^0+p^0)}-1}\right)} \bbGA(\zeta_2|\zeta_1,p) \ .
    \end{split}
\end{equation}
Here $\bbGR(\zeta_1|\zeta_1,p)$ and $\bbGA(\zeta_1|\zeta_1,p)$ are the retarded and the advanced bulk-to-bulk propagators respectively, given in Eq.~\eqref{eq:DefScalarBlkBlkRetAdv}. From the above result, note that the temperature-dependent factors outside the propagators are all either Fermi-Dirac or Bose-Einstein statistical factors, depending on the parity of $\alpha_1$ and $\alpha_2$. In particular, note that for $\alpha_1=\alpha_2=0$, we obtain only Bose-Einstein factors and that the result matches the purely scalar theory result in \cite{Loganayagam:2024mnj}.

We now come to the double-discontinuity involving the spinor bulk-to-bulk propagator. Before proceeding to perform the monodromy integrals, we take a moment to note that exactly as in the purely scalar case, here too there are no pole contributions from the horizon. This might seem surprising in the case of fermions since there are factors of $1/\sqrt{f}$ that can multiply to give horizon poles of the form $1/f$. This does not happen because the factors of $1/\sqrt{f}$ in all our computations only come from the outgoing propagator. In particular, from the time-reversal matrix. Since the time-reversal matrix is idempotent (squares to identity), there can never be factors of $1/f$ in the calculation, thus avoiding the question of horizon-pole contributions.

Accounting for the monodromy across the branch cut, the double-discontinuity of the spinor bulk-to-bulk propagator takes the form 
\begin{equation}
    \begin{split}
        \mathscr{I}_{\bbS}= \int_{\text{ext}_1} \int_{\text{ext}_2}  \ [f(\zeta_1)]^{\frac{\alpha_1}{2}} [f(\zeta_2)]^{\frac{\alpha_2}{2}}e^{-\beta k^0_1 \zeta_1} e^{-\beta k^0_2 \zeta_2} \bar{\mathfrak{a}}_2(\zeta_2)\bbSDD(\zeta_2|\zeta_1,p) \bar{\mathfrak{a}}_1(\zeta_1) \ ,
    \end{split}
\end{equation}
where $\bbSDD(\zeta_2|\zeta_1,p)$ is the double discontinuity
\begin{equation}
    \begin{split}
        \bbSDD(\zeta_2|\zeta_1,p) &\equiv \bbS(\zeta_2|\zeta_1,p) - (-)^{\alpha_1} e^{-\beta k^0_1 }  \bbS(\zeta_2|\zeta_1 +1,p)\\
        &  - (-)^{\alpha_2} e^{-\beta k^0_2 }  \bbS(\zeta_2+1|\zeta_1,p)+ (-)^{\alpha_1+\alpha_2} e^{-\beta k^0_1 } e^{-\beta k^0_2 } \bbS(\zeta_2+1|\zeta_1+1,p) \ .
    \end{split}
\end{equation}
This expression can now be greatly simplified using the theta function identities
\begin{equation}
    \begin{split}
        &\thetaSK(\zeta_1 + 1 > \zeta_2 +1) = \thetaSK(\zeta_1 >\zeta_2)  \ ,\\
        & \thetaSK(\zeta_1+1 > \zeta_2) = 1\ , \qquad  \thetaSK(\zeta_1 > \zeta_2 +1 ) = 0 \ .
    \end{split}
\end{equation}
These hold when $\zeta_1$ and $\zeta_2$ both belong to the upper branch of the grSK contour, which is indeed the case here. With the explicit form of the bulk-to-bulk propagator as well, the double-discontinuity can be written as
\begin{equation}
    \begin{split}
        \bbSDD(\zeta_2|\zeta_1,p) &= \Sin (\zeta_2,p)\frac{1}{\left\{\Kret(p)+\Kret(-p) \right\}} \left[ c_1 \Sbin(\zeta_1,-p)-c_2 \Sbout(\zeta_1,-p) \right] \\
        &+\Sout (\zeta_2,p)\frac{1}{\left\{\Kret(p)+\Kret(-p) \right\}} \left[ c_3 \Sbin(\zeta_1,-p)-c_4 \Sbout(\zeta_1,-p) \right]  \ ,
    \end{split}
\end{equation}
where
\begin{equation}
    \begin{split}
        c_2 &= n^{\fd}_{p^0}\left(e^{i \alpha_1 \pi} e^{-\beta (k_1^0-p^0)}+1\right)\left(e^{i \alpha_2 \pi} e^{-\beta k_2^0}-1\right) \ ,\\
        c_3 &= (1-n^{\fd}_{p^0}) \left(e^{i \alpha_1 \pi} e^{-\beta k_1^0}-1 \right) \left(e^{i \alpha_2 \pi} e^{-\beta (k_2^0 +p^0)}+1 \right) \ ,\\
        c_1 &= (1+c_2)\thetaSK(\zeta_2<\zeta_1) - (1+c_3) \thetaSK(\zeta_2>\zeta_1)\ ,\\ 
        c_4 &= (1+c_3)\thetaSK(\zeta_2<\zeta_1) - (1+c_2) \thetaSK(\zeta_2>\zeta_1)\ .
    \end{split}
\end{equation}

Notice that the factors in $c_2$ and $c_3$ are just Fermi-Dirac or Bose-Einstein statistical factors depending on the parity of $\alpha_1$ and $\alpha_2$. Moreover, there are only two independent combinations of statistical factors here. Thus we can indeed rewrite the above expression as
\begin{equation}
    \begin{split}
        \bbSDD(\zeta_2|\zeta_1,p) &= -\left(e^{i \alpha_1 \pi} e^{-\beta (k^0_1-p^0)}+1\right)\left(e^{i \alpha_2 \pi} e^{-\beta k^0_2}-1\right) n^{\fd}_{p^0} \bbSR(\zeta_2|\zeta_1,p)\\
        &- \left(e^{i \alpha_1 \pi} e^{-\beta k^0_1}-1 \right) \left(e^{i \alpha_2 \pi} e^{-\beta (k^0_2 +p^0)}+1 \right) (1-n^{\fd}_{p^0}) \bbSA(\zeta_2|\zeta_1,p)\ ,
    \end{split}
    \label{eq:DoubleDiscSpinor}
\end{equation}
where $\bbSR(\zeta_2|\zeta_1,p)$ and $\bbSA(\zeta_2|\zeta_1,p)$ are the retarded and the advanced bulk-to-bulk propagators given in Eq.~\eqref{eq:DefRetardedSpinorBlkBlk} and Eq.~\eqref{eq:DefAdvancedSpinorBlkBlk} respectively.

So far, we have focused on evaluating double-discontinuity integrals relevant to the computation of the four-point exchange influence phase. To evaluate higher point functions, we require higher monodromy integrals with richer structures of products of bulk-to-bulk propagators. For example, at the level of five-point functions, we require integrals of the form
\begin{equation}
    \oint_{\zeta_1} \oint_{\zeta_2} \oint_{\zeta_3} \ [f(\zeta_1)]^{\frac{\alpha_1}{2}} [f(\zeta_2)]^{\frac{\alpha_2}{2}}  [f(\zeta_3)]^{\frac{\alpha_3}{2}}  e^{-\beta (k^0_1  \zeta_1+k^0_2  \zeta_2+k^0_3  \zeta_3)}    \bbG(\zeta_2|\zeta_1, p_1)  \bar{\mathfrak{a}}_2(\zeta_3, \zeta_1)\bbS(\zeta_{3}|\zeta_{2}, p_2) \mathfrak{a}_1(\zeta_2)
\end{equation}
and
\begin{equation}
    \oint_{\zeta_1} \oint_{\zeta_2} \oint_{\zeta_3} \ [f(\zeta_1)]^{\frac{\alpha_1}{2}} [f(\zeta_2)]^{\frac{\alpha_2}{2}}  [f(\zeta_3)]^{\frac{\alpha_3}{2}} e^{-\beta (k^0_1  \zeta_1+k^0_2  \zeta_2+k^0_3  \zeta_3)}  \bar{\mathfrak{a}}_2(\zeta_3, \zeta_2) \bbS(\zeta_{3}|\zeta_{2}, p_2) \bbS(\zeta_{2}|\zeta_{1}, p_1) \mathfrak{a}_1(\zeta_1)  \ .
\end{equation}
These integrals can also be computed by the techniques presented above and can also be arranged in terms of the retarded and the advanced propagators for scalars and spinors, presented above. Since we are only interested in the four-point functions, we do not present these results here.

\section{Formal perturbative expansion of the influence phase}\label{sec:AppendixPertExpInfPhase}

In this appendix, we will present a formal perturbative expansion of the Yukawa influence phase in terms of the free solutions and the bulk-to-bulk propagators. 

The Euler-Lagrange equations obtained by varying the Yukawa action in Eq.~\eqref{eq:ActionYukawa} are
\begin{equation}
    \Box \phi = i\lambda \bar{\Psi}  \  \qquad \text{and} \qquad \Psi \Gamma^{\A} D_{\A} \Psi = \lambda \phi \Psi \ ,
\end{equation}
where $\Box$ is the wave operator of the AdS$_{d+1}$ Schwarzschild blackbrane, and $D_{\A}$ is the covariant derivative acting on the spinor fields defined in Eq.~\eqref{eq:CovDerDiracSpinors}. These equations can now be solved perturbatively in the Yukawa coupling $\lambda$. In other words, we will write the solutions in the form given in Eq.~\eqref{eq:FormalPertExpPsi} and Eq.~\eqref{eq:FormalPertExpphi}. The leading-order solutions $\Psi_{(0)}$ and $\phi_{(0)}$ solve the free Dirac equation and the free wave equation respectively. The higher-order solutions can be found by solving the equations of motion above perturbatively. 

We take the leading-order solutions to satisfy the GKPW boundary conditions, and thus the higher-order solutions must all be normalisable on both the boundaries. Therefore, we can write expressions of the form
\begin{equation}
    \Psi_{(n)} (\zeta,k) = \oint_{\zeta} \ \bbS(\zeta|\zeta_0,k)  \mathfrak{J}_{(n)} (\zeta_0,k)\ , \quad \text{and} \quad \phi_{(n)} (\zeta,k) = \oint_{\zeta} \ \bbG(\zeta|\zeta_0,k)  \mathbb{J}_{(n)} (\zeta_0,k) \ ,
    \label{eq:Psi1andphi1}
\end{equation}
where $\bbS(\zeta|\zeta_0,k)$ and $\bbG(\zeta|\zeta_0,k)$ are the spinor and scalar bulk-to-propagators, which are normalisable on both the boundaries, respectively. Here $\mathfrak{J}_{(n)} (\zeta_0,k)$ and $\mathbb{J}_{(n)} (\zeta_0,k)$ are sources for the $n$th order solutions, which can be read-off from the equations of motion. For example, at the first and the second orders, we have the expressions given in Eq.~\eqref{eq:BulkSources1} and Eq.~\eqref{eq:BulkSources2} respectively.

The above perturbative expansions of the solutions can now be substituted into the action to provide the on-shell action, which is the influence phase at leading order.\footnote{Here, leading order refers to the leading term in $1/N$ expansion in the boundary theory.} We will arrange this expansion with the notation given in Eq.~\eqref{eq:OnShellActionPertExp}.
The symbol $S_{(2)}$ denotes the part of the on-shell action that is quadratic in the boundary sources and takes the form
\begin{equation}
    S_{(2)} = -\frac{1}{2} \int \diff \Sigma^{\A} \phi_{(0)} \partial_{\A} \phi_{(0)} +\int \diff^d \bx \ r^{d} \left(i \overline{\Psi}_{(0)} \mathbb{P}_- \Psi_{(0)}\right)\Big|_{\zeta = 0}^{\zeta =1} \ .
    \label{eq:ActionS2}
\end{equation}

The term linear in the interaction parameter $\lambda$ is also cubic in the boundary sources and takes the form
\begin{equation}
    \begin{split}
        S_{(3)} &= \oint \diff \zeta \int \diff^d \bx \sqrt{-g} \  \overline{\Psi}_{(0)} i \Gamma^{\A} D_{\A} \Psi_{(1)} - \oint \diff \zeta \int \diff^d \bx \sqrt{-g} \  \partial_{\A} \phi_{(0)} \partial^{\A} \phi_{(1)}\\
        &+ \int \diff^d \bx \ r^{d} \left[i \overline{\Psi}_{(0)} \mathbb{P}_- \Psi_{(1)}+i \overline{\Psi}_{(1)} \mathbb{P}_- \Psi_{(0)}\right]\Big|_{\zeta = 0}^{\zeta =1}- i \oint \diff \zeta \int \diff^d \bx \sqrt{-g} \  \phi_{(0)} \overline{\Psi}_{(0)} \Psi_{(0)} \ .
    \end{split}
\end{equation}
This expression can be greatly simplified using the equations of motion satisfied by the fields. Firstly, note that the first-order corrections (in the solutions) satisfy Eq.~\eqref{eq:Psinandphin} with $n=1$. Using the explicit form of the $n=1$ sources in Eq.~\eqref{eq:BulkSources1}, we see that the first term cancels the final term. The second term vanishes, as can be seen by using integration by parts and noting that the first-order solution vanishes at the boundary. This leaves only the boundary term in the square brackets. Again, the second term here vanishes because of the boundary conditions on $\Psibar_{(1)}$. Thus, we finally get only one simple term
\begin{equation}
    \begin{split}
        S_{(3)} &= \int \diff^d \bx \ r^{d} \ i \overline{\Psi}_{(0)} \mathbb{P}_- \Psi_{(1)}\Big|_{\zeta = 0}^{\zeta =1} = \ i r^d \int \frac{\diff ^d k}{(2 \pi)^d} \Psibar_{(0)}(-k) \projm \Psi_{(1)} (k) \Big|_{\zeta = 0}^{\zeta= 1}\ .
    \end{split}
    \label{eq:ThreePtInfPhaseFormalExprFinal}
\end{equation}
We now obtain an explicit expression for $S_{(3)}$ in terms of the free solutions by using the first-order solutions, and simplifying using the boundary limits of the bulk-to-bulk propagator (Eq.~\eqref{eq:BoundaryLimitsSpinorBlkBlk}) to get
\begin{equation}
    \begin{split}
        S_{(3)} &= i r^{d} \int_{k_1}\int_{k} (2\pi)^{d} \delta^{(d)} (k_1+k)  \overline{\Psi}_{(0)}(\zeta,k_1) \mathbb{P}_- \oint_{\zeta_0} \bbS(\zeta|\zeta_0,k) \mathfrak{J}_{(1)}(\zeta_0,k)\Big|_{\zeta = 0}^{\zeta =1}\\
        &= -i  \int_{k_{1,2,3}} \oint_{\zeta_0}   \bigg[\overline{\psi}_{\sR}(k_1)\bar{s}_{\sR}(\zeta_0,k_1)- \overline{\psi}_{\sL}(k_1)   \bar{s}_{\sL}(\zeta_0,k_1) \bigg]  \phi_{(0)}(\zeta_0,k_2) \Psi_{(0)}(\zeta_0,k_3) \\
        &= -i \int_{k_{1,2,3}} \oint_{\zeta} \Psibar_{(0)}(\zeta,k_1)    \phi_{(0)}(\zeta,k_2) \Psi_{(0)}(\zeta,k_3) \ ,
    \end{split}
    \label{eq:ThreePtInfPhaseintermsofSonTeaneySols}
\end{equation}
where we have employed the notation for momentum integrals introduced in Eq.~\eqref{eq:DefMomentumIntegral}.

The correction that is quadratic in $\lambda$ or quartic in the boundary sources takes the form
\begin{equation}
    S_{(4)} = \frac{i}{2} \oint \diff \zeta \diff^d \bx \sqrt{-g} \ \phi_{(1)} \Psibar_{(0)} \Psi_{(0)}+i r^d \int \frac{\diff ^d k}{(2 \pi)^d} \Psibar_{(0)}(-k) \projm \Psi_{(2)} (k) \Big|_{\zeta = 0}^{\zeta= 1}\ .
\end{equation}
Using the bulk sources in Eq.~\eqref{eq:BulkSources1} and Eq.~\eqref{eq:BulkSources2}, we obtain
\begin{equation}
    \begin{split}
        S_{(4)} =   &\red{-}\frac{1}{2} \int_{k_{1,2,3,4}} \oint_{\z_1} \oint_{\z_2}  \overline{\Psi}_{(0)}(\z_2,k_1) \Psi_{(0)}(\z_2,k_2) \bbG(\z_2 |\z_1,k_3 +k_4) \overline{\Psi}_{(0)}(\z_1,k_3) \Psi_{(0)}(\z_1,k_4)\\
        &+i \int_{k_{1,2,3,4}} \oint_{\z_1} \oint_{\z_2} \phi_{(0)}(\z_2,k_1)  \overline{\Psi}_{(0)}(\z_2,k_2)  \bbS(\z_2 |\z_1,k_3 +k_4) \phi_{(0)}(\z_1,k_3) \Psi_{(0)}(\z_1,k_4) \ .
    \end{split}
    \label{eq:FourPtInfPhaseintermsofSonTeaneySols}
\end{equation}

\section{Four-point influence phase in the past-future basis}\label{sec:AppendixFourPointcolumnVector}
In this appendix, we will give the explicit results for all the terms present in the four-point Yukawa influence phase.

\subsection*{Terms with bulk scalar exchange}
In what follows, we will use the notation introduced in Eq.~\eqref{eq:DefBraketNotationSpectralFunc} to avoid clutter. The terms in the column vector representation are:
{\small
\begin{equation}
    \begin{split}
        &S_{\Fb \Fb \Fb \Pb,\bbG} =  \int_{k_{1,2,3,4}} \Big\{ \langle \Sbin \Sin \bbGA \Sbin  \Sout \rangle \cdot   \fF^{{}_{(1)}}(\bar{k}_1)  \otimes \fF^{{}_{(1)}}(\bar{k}_2) \otimes \fF^{{}_{(1)}}(\bar{k}_3)  \otimes \left[ \fP^{{}_{(2)}}(\bar{k}_4)-\fP^{{}_{(1)}}(\bar{k}_4)\right]\\
        &\hspace{2cm}+\langle  \Sbin \Sin \bbGA \Sbout  \Sin  \rangle \cdot  \fF^{{}_{(1)}}(\bar{k}_1)  \otimes \fF^{{}_{(1)}}(\bar{k}_2) \otimes \left[ \fP^{{}_{(2)}}(\bar{k}_3)-\fP^{{}_{(1)}}(\bar{k}_3)\right]  \otimes \fF^{{}_{(1)}}(\bar{k}_4) \Big\} \ ,
    \end{split}
\end{equation}
}

{\small
\begin{equation}
    \begin{split}
        &S_{\Fb \Pb \Pb \Pb,\bbG}= \int_{k_{1,2,3,4}}\\
        &\times\Big\{ \langle \Sbin \Sout \bbGA \Sbout  \Sout \rangle \cdot  \fF^{{}_{(1)}}(\bar{k}_1) \otimes \left[ \fP^{{}_{(2)}}(\bar{k}_2) \otimes \fP^{{}_{(2)}}(\bar{k}_3) \otimes \fP^{{}_{(2)}}(\bar{k}_4)-\fP^{{}_{(1)}}(\bar{k}_2) \otimes \fP^{{}_{(1)}}(\bar{k}_3) \otimes \fP^{{}_{(1)}}(\bar{k}_4)\right]  \\
        &+\langle  \Sbout \Sin \bbGA \Sbout  \Sout  \rangle \cdot  \fF^{{}_{(1)}}(\bar{k}_2) \otimes \left[ \fP^{{}_{(2)}}(\bar{k}_1) \otimes \fP^{{}_{(2)}}(\bar{k}_3) \otimes \fP^{{}_{(2)}}(\bar{k}_4)-\fP^{{}_{(1)}}(\bar{k}_1) \otimes \fP^{{}_{(1)}}(\bar{k}_3) \otimes \fP^{{}_{(1)}}(\bar{k}_4)\right] \Big\}\ ,
    \end{split}
\end{equation}
}

{\small
\begin{equation}
    \begin{split}
        &S_{\Fb \Fb \Pb \Pb,\bbG} = \red{-}\int_{k_{1,2,3,4}} \langle \Sbin \Sin \bbGA \Sbout  \Sout \rangle \cdot  \fF^{{}_{(1)}}(\bar{k}_1) \otimes \fF^{{}_{(1)}}(\bar{k}_2) \otimes  \left[ \fP^{{}_{(2)}}(\bar{k}_3) \otimes \fP^{{}_{(2)}}(\bar{k}_4) -\fP^{{}_{(1)}}(\bar{k}_3) \otimes \fP^{{}_{(1)}}(\bar{k}_4)\right]    \ ,
    \end{split}
\end{equation}
}
and
\begin{equation}
    \begin{split}
        &S_{\Fb \Pb \Fb \Pb,\bbG} = \red{-}\int_{k_{1,2,3,4}} \Big\{  \bar{\rho}^{(1)}_{\bbG}[21][43]  \cdot \fF^{{}_{(1)}}(\bar{k}_1) \otimes \fP^{{}_{(2)}}(\bar{k}_2) \otimes \fF^{{}_{(1)}}(\bar{k}_3)  \otimes  \left[ \fP^{{}_{(2)}}(\bar{k}_4)- \fP^{{}_{(1)}}(\bar{k}_4)\right] \\
        &\hspace{3cm}  - \bar{\rho}^{(1)}_{\bbG}[43][21]  \cdot \fF^{{}_{(1)}}(\bar{k}_1) \otimes \fP^{{}_{(1)}}(\bar{k}_2) \otimes \fF^{{}_{(1)}}(\bar{k}_3)  \otimes  \left[ \fP^{{}_{(2)}}(\bar{k}_4)- \fP^{{}_{(1)}}(\bar{k}_4)\right]  \\
        &\hspace{3cm}+ \bar{\rho}^{(2)}_{\bbG}[12][43]  \cdot \fP^{{}_{(2)}}(\bar{k}_1) \otimes \fF^{{}_{(1)}}(\bar{k}_2) \otimes \fF^{{}_{(1)}}(\bar{k}_3)  \otimes  \left[ \fP^{{}_{(2)}}(\bar{k}_4)- \fP^{{}_{(1)}}(\bar{k}_4)\right] \\
        &\hspace{3cm}  - \bar{\rho}^{(2)}_{\bbG}[43][12]  \cdot \fP^{{}_{(1)}}(\bar{k}_1) \otimes \fF^{{}_{(1)}}(\bar{k}_2) \otimes \fF^{{}_{(1)}}(\bar{k}_3)  \otimes  \left[ \fP^{{}_{(2)}}(\bar{k}_4)- \fP^{{}_{(1)}}(\bar{k}_4)\right]\\
        &\hspace{3cm}+ \bar{\rho}^{(3)}_{\bbG}[21][34]  \cdot \fF^{{}_{(1)}}(\bar{k}_1) \otimes \fP^{{}_{(2)}}(\bar{k}_2)    \otimes  \left[ \fP^{{}_{(2)}}(\bar{k}_3)- \fP^{{}_{(1)}}(\bar{k}_3)\right]  \otimes \fF^{{}_{(1)}}(\bar{k}_4) \\
        &\hspace{3cm}  - \bar{\rho}^{(3)}_{\bbG}[34][21]  \cdot \fF^{{}_{(1)}}(\bar{k}_1) \otimes \fP^{{}_{(1)}}(\bar{k}_2)    \otimes  \left[ \fP^{{}_{(2)}}(\bar{k}_3)- \fP^{{}_{(1)}}(\bar{k}_3)\right]  \otimes \fF^{{}_{(1)}}(\bar{k}_4) \\  &\hspace{3cm}+ \bar{\rho}^{(4)}_{\bbG}[12][34]  \cdot \fP^{{}_{(2)}}(\bar{k}_1) \otimes \fF^{{}_{(1)}}(\bar{k}_2)    \otimes  \left[ \fP^{{}_{(2)}}(\bar{k}_3)- \fP^{{}_{(1)}}(\bar{k}_3)\right]  \otimes \fF^{{}_{(1)}}(\bar{k}_4) \\
        &\hspace{3cm}  - \bar{\rho}^{(4)}_{\bbG}[34][12]  \cdot \fP^{{}_{(1)}}(\bar{k}_1) \otimes \fF^{{}_{(1)}}(\bar{k}_2)    \otimes  \left[ \fP^{{}_{(2)}}(\bar{k}_3)- \fP^{{}_{(1)}}(\bar{k}_3)\right]  \otimes \fF^{{}_{(1)}}(\bar{k}_4)\Big\} \ , 
    \end{split}
\end{equation}
where we have defined
\begin{equation}
    \begin{split}
       \bar{\rho}^{(1)}_{\bbG}[43][21] &\equiv  \  
        n^{\be}_{k^{0}_3+k^0_4}\langle \Sbin \Sout \bbGA \Sbin  \Sout \rangle  \ , \quad \bar{\rho}^{(1)}_{\bbG}[21][43] \equiv  e^{-\beta k^0_2 -\beta k^0_1} \bar{\rho}^{(1)}_{\bbG}[43][21] \ ,\\
       \bar{\rho}^{(2)}_{\bbG}[43][12] &\equiv  \  
        n^{\be}_{k^{0}_3+k^0_4}\langle \Sbout \Sin \bbGA \Sbin  \Sout \rangle \ , \quad  \bar{\rho}^{(2)}_{\bbG}[12][43]  = e^{-\beta k^0_2 -\beta k^0_1} \bar{\rho}^{(2)}_{\bbG}[43][12] \ ,\\
       \bar{\rho}^{(3)}_{\bbG}[34][21] &\equiv  \  
        n^{\be}_{k^{0}_3+k^0_4}\langle \Sbin \Sout \bbGA \Sbout  \Sin \rangle  \ , \quad \bar{\rho}^{(3)}_{\bbG}[21][34] = e^{-\beta k^0_2 -\beta k^0_1} \bar{\rho}^{(3)}_{\bbG}[34][21]\ ,\\
        \bar{\rho}^{(4)}_{\bbG}[34][12] &\equiv  \  
        n^{\be}_{k^{0}_3+k^0_4}  \langle \Sbout \Sin \bbGA \Sbout  \Sin \rangle \ , \quad  \bar{\rho}^{(4)}_{\bbG}[12][34]  = e^{-\beta k^0_2 -\beta k^0_1} \bar{\rho}^{(4)}_{\bbG}[34][12]\ .
    \end{split} 
\end{equation}

\subsection*{Terms with bulk spinor exchange}
Again for notational convenience, as we did in the case of bulk scalar exchange, we define
{\small
\begin{equation}
    \begin{split}
        \langle G^{x_1} \overline{S}^{x_2}  \bbS_{x}  G^{x_3} S^{x_4}  \rangle  &\equiv \int_{\text{ext}_1} \int_{\text{ext}_2} G^{x_1}(\zeta_2, k_1) \left[\overline{S}^{x_2}(\zeta_2, k_2)\right]^{\alpha}_{\ \bulkalpha}  \left[\bbS_{x}(\zeta_2|\z_1, k_3+k_4)\right]^{\bulkalpha}_{\ \bulkbeta}\\
        &\hspace{2cm} \times   G^{x_3}(\zeta_1, k_3)  \left[S^{x_4}(\zeta_1, k_4)\right]^{\bulkbeta}_{\ \delta} \boldsymbol{J}(k_1) \otimes \left[  \boldsymbol{\psibar}(k_2)  \right]_{\alpha}  \otimes     \boldsymbol{J}(k_3) \otimes  \left[ \boldsymbol{\psi}(k_4)  \right]^{\beta}   \ ,      
    \end{split}
\end{equation}
}
where $x_1, x_2, x_3, x_4 \in \{\text{in}, \text{out}\}$ and $x \in \{\text{ret}, \text{adv}\}$. Then, all the terms in the column vector representation are:
{\small
\begin{equation}
    \begin{split}
        S_{\Fb \Fb \Fb \Pb,\bbS}        &= \ \red{-}i  \int_{k_{1,2,3,4}}\Big\{ \langle \Gin \Sbin   \bbSA  \Gin \Sout \rangle  \cdot \eF^{{}_{(1)}}(\bar{k}_1)   \otimes \fF^{{}_{(1)}}(\bar{k}_3) \otimes \eF^{{}_{(1)}}(\bar{k}_3) \otimes \left[ \fP^{{}_{(2)}}(\bar{k}_4)-\fP^{{}_{(1)}}(\bar{k}_4)\right]\\
        &\hspace{1.8cm}+ \langle \Gin \Sbin   \bbSA  \Gout \Sin \rangle  \cdot  \eF^{{}_{(1)}}(\bar{k}_1) \otimes \fF^{{}_{(1)}}(\bar{k}_2) \otimes \left[ \eP^{{}_{(2)}}(\bar{k}_3)-\eP^{{}_{(1)}}(\bar{k}_3)\right]\otimes \fF^{{}_{(1)}}(\bar{k}_4) \\
        &\hspace{1.8cm}+ \langle  \Gin \Sbout   \bbSR  \Gin \Sin \rangle  \cdot  \eF^{{}_{(1)}}(\bar{k}_1) \otimes \left[ \fP^{{}_{(2)}}(\bar{k}_2) - \fP^{{}_{(1)}}(\bar{k}_2) \right] \otimes \eF^{{}_{(1)}}(\bar{k}_3)\otimes \fF^{{}_{(1)}}(\bar{k}_4) \\
        &\hspace{1.8cm}+ \langle \Gout \Sbin   \bbSR  \Gin \Sin \rangle   \cdot  \left[ \eP^{{}_{(2)}}(\bar{k}_1)-\eP^{{}_{(1)}}(\bar{k}_1)\right] \otimes  \fF^{{}_{(1)}}(\bar{k}_2) \otimes \eF^{{}_{(1)}}(\bar{k}_3) \otimes \fF^{{}_{(1)}}(\bar{k}_4) \Big\} \ ,
    \end{split}
\end{equation}
}

{\small
\begin{equation}
    \begin{split}
        &S_{\Fb \Pb \Pb \Pb,\bbS}= \red{-} i  \int_{k_{1,2,3,4}}\\
        & \times \Big\{ \langle \Gin \Sbout   \bbSA  \Gout \Sout \rangle \cdot    \eF^{(1)}(\bar{k}_1) \otimes \left( \fP^{(2)}(\bar{k}_2) \otimes \eP^{(2)}(\bar{k}_3) \otimes  \fP^{(2)}(\bar{k}_4)-\fP^{(1)}(\bar{k}_2) \otimes \eP^{(1)}(\bar{k}_3) \otimes \fP^{(1)}(\bar{k}_4)\right)\\
        &+\langle \Gout \Sbin   \bbSA  \Gout \Sout \rangle \cdot  \fF^{(1)}(\bar{k}_2) \otimes \left( \eP^{(2)}(\bar{k}_1) \otimes \eP^{(2)}(\bar{k}_3) \otimes  \fP^{(2)}(\bar{k}_4)-\eP^{(1)}(\bar{k}_1) \otimes \eP^{(1)}(\bar{k}_3) \otimes \fP^{(1)}(\bar{k}_4)\right)\\
        &+\langle \Gout \Sbout   \bbSR  \Gin \Sout \rangle \cdot \eF^{(1)}(\bar{k}_3) \otimes \left( \eP^{(2)}(\bar{k}_1) \otimes \fP^{(2)}(\bar{k}_2) \otimes  \fP^{(2)}(\bar{k}_4)-\eP^{(1)}(\bar{k}_1) \otimes \fP^{(1)}(\bar{k}_2) \otimes \fP^{(1)}(\bar{k}_4)\right)\\
        &+\langle \Gout \Sbout   \bbSR  \Gout \Sin \rangle \cdot  \left( \eP^{(2)}(\bar{k}_1) \otimes \fP^{(2)}(\bar{k}_2) \otimes \eP^{(2)}(\bar{k}_3) -\eP^{(1)}(\bar{k}_1) \otimes \fP^{(1)}(\bar{k}_2) \otimes \eP^{(1)}(\bar{k}_3) \right)\otimes  \fF^{(1)}(\bar{k}_4)\Big\}  \ ,
\end{split}
\end{equation}
}

{\small
\begin{equation}
    \begin{split}
        S_{ \Fb \Fb \Pb \Pb,\bbS}  &= \ i  \int_{k_{1,2,3,4}}\\
        &\times \Big\{ \langle \Gin \Sbin   \bbSA  \Gout \Sout \rangle \cdot    \eF^{(1)}(\bar{k}_1) \otimes  \fF^{(1)}(\bar{k}_2) \otimes\left(  \eP^{(2)}(\bar{k}_3) \otimes  \fP^{(2)}(\bar{k}_4)- \eP^{(1)}(\bar{k}_3) \otimes \fP^{(1)}(\bar{k}_4)\right)\\
        &+\langle \Gout \Sbout   \bbSR  \Gin \Sin \rangle \cdot   \left(  \eP^{(2)}(\bar{k}_1) \otimes  \fP^{(2)}(\bar{k}_2)- \eP^{(1)}(\bar{k}_1) \otimes \fP^{(1)}(\bar{k}_2)\right) \otimes \eF^{(1)}(\bar{k}_3) \otimes  \fF^{(1)}(\bar{k}_4)\Big\} \ ,
\end{split}
\end{equation}
}
and
\begin{equation}
    \begin{split}
        &S_{\Fb \Pb \Fb \Pb,\bbS} = \red{-}i \int_{k_{1,2,3,4}} \Big\{  \bar{\rho}^{(1)}_{\bbS}[21][43]  \cdot  \eF^{{}_{(1)}}(\bar{k}_1) \otimes \fP^{{}_{(2)}}(\bar{k}_2) \otimes \eF^{{}_{(1)}}(\bar{k}_3) \otimes  \left[ \fP^{{}_{(2)}}(\bar{k}_4) -\fP^{{}_{(1)}}(\bar{k}_4) \right] \\
        &\hspace{3cm}  - \bar{\rho}^{(1)}_{\bbS}[43][21]  \cdot  \eF^{{}_{(1)}}(\bar{k}_1) \otimes \fP^{{}_{(1)}}(\bar{k}_2) \otimes \eF^{{}_{(1)}}(\bar{k}_3) \otimes  \left[ \fP^{{}_{(2)}}(\bar{k}_4) -\fP^{{}_{(1)}}(\bar{k}_4) \right] \\
        &\hspace{3cm} + \bar{\rho}^{(2)}_{\bbS}[43][21]  \cdot   \eF^{{}_{(1)}}(\bar{k}_1) \otimes \left[ \fP^{{}_{(2)}}(\bar{k}_2) -\fP^{{}_{(1)}}(\bar{k}_2) \right] \otimes \eF^{{}_{(1)}}(\bar{k}_3) \otimes  \fP^{{}_{(2)}}(\bar{k}_4) \\
        &\hspace{3cm}  - \bar{\rho}^{(2)}_{\bbS}[21][43]  \cdot   \eF^{{}_{(1)}}(\bar{k}_1) \otimes \left[ \fP^{{}_{(2)}}(\bar{k}_2) -\fP^{{}_{(1)}}(\bar{k}_2) \right] \otimes \eF^{{}_{(1)}}(\bar{k}_3) \otimes  \fP^{{}_{(1)}}(\bar{k}_4)\\
        &\hspace{3cm}+ \bar{\rho}^{(3)}_{\bbS}[12][43]  \cdot  \eP^{{}_{(2)}}(\bar{k}_1) \otimes \fF^{{}_{(1)}}(\bar{k}_2) \otimes \eF^{{}_{(1)}}(\bar{k}_3) \otimes  \left[ \fP^{{}_{(2)}}(\bar{k}_4) -\fP^{{}_{(1)}}(\bar{k}_4) \right] \\
        &\hspace{3cm}  - \bar{\rho}^{(3)}_{\bbS}[43][12]  \cdot  \eP^{{}_{(1)}}(\bar{k}_1) \otimes \fF^{{}_{(1)}}(\bar{k}_2) \otimes \eF^{{}_{(1)}}(\bar{k}_3) \otimes  \left[ \fP^{{}_{(2)}}(\bar{k}_4) -\fP^{{}_{(1)}}(\bar{k}_4) \right]\\
        &\hspace{3cm}+ \bar{\rho}^{(4)}_{\bbS}[43][12]  \cdot \left[  \eP^{{}_{(2)}}(\bar{k}_1)- \eP^{{}_{(1)}}(\bar{k}_1)\right]  \otimes \fF^{{}_{(1)}}(\bar{k}_2) \otimes \eF^{{}_{(1)}}(\bar{k}_3) \otimes \fP^{{}_{(2)}}(\bar{k}_4) \\
        &\hspace{3cm}  - \bar{\rho}^{(4)}_{\bbS}[12][43]  \cdot \left[  \eP^{{}_{(2)}}(\bar{k}_1)- \eP^{{}_{(1)}}(\bar{k}_1)\right]  \otimes \fF^{{}_{(1)}}(\bar{k}_2) \otimes \eF^{{}_{(1)}}(\bar{k}_3) \otimes \fP^{{}_{(1)}}(\bar{k}_4)\\
        &\hspace{3cm}+ \bar{\rho}^{(5)}_{\bbS}[21][34]  \cdot  \eF^{{}_{(1)}}(\bar{k}_1) \otimes \fP^{{}_{(2)}}(\bar{k}_2) \otimes \left[ \eP^{{}_{(2)}}(\bar{k}_3)-\eP^{{}_{(1)}}(\bar{k}_3)\right] \otimes \fF^{{}_{(1)}}(\bar{k}_4) \\
        &\hspace{3cm}  - \bar{\rho}^{(5)}_{\bbS}[34][21]  \cdot  \eF^{{}_{(1)}}(\bar{k}_1) \otimes \fP^{{}_{(1)}}(\bar{k}_2) \otimes \left[ \eP^{{}_{(2)}}(\bar{k}_3)-\eP^{{}_{(1)}}(\bar{k}_3)\right] \otimes \fF^{{}_{(1)}}(\bar{k}_4) \\
        &\hspace{3cm} + \bar{\rho}^{(6)}_{\bbS}[34][21]  \cdot   \eF^{{}_{(1)}}(\bar{k}_1) \otimes \left[ \fP^{{}_{(2)}}(\bar{k}_2) -\fP^{{}_{(1)}}(\bar{k}_2) \right] \otimes \eP^{{}_{(2)}}(\bar{k}_3) \otimes  \fF^{{}_{(1)}}(\bar{k}_4) \\
        &\hspace{3cm}  - \bar{\rho}^{(6)}_{\bbS}[21][34]  \cdot   \eF^{{}_{(1)}}(\bar{k}_1) \otimes \left[ \fP^{{}_{(2)}}(\bar{k}_2) -\fP^{{}_{(1)}}(\bar{k}_2) \right] \otimes \eP^{{}_{(1)}}(\bar{k}_3) \otimes  \fF^{{}_{(1)}}(\bar{k}_4) \\
        &\hspace{3cm}+ \bar{\rho}^{(7)}_{\bbS}[12][34]  \cdot  \eP^{{}_{(2)}}(\bar{k}_1) \otimes \fF^{{}_{(1)}}(\bar{k}_2) \otimes \left[ \eP^{{}_{(2)}}(\bar{k}_3)-\eP^{{}_{(1)}}(\bar{k}_3)\right] \otimes \fF^{{}_{(1)}}(\bar{k}_4) \\
        &\hspace{3cm}  - \bar{\rho}^{(7)}_{\bbS}[34][12]  \cdot  \eP^{{}_{(1)}}(\bar{k}_1) \otimes \fF^{{}_{(1)}}(\bar{k}_2) \otimes \left[ \eP^{{}_{(2)}}(\bar{k}_3)-\eP^{{}_{(1)}}(\bar{k}_3)\right] \otimes \fF^{{}_{(1)}}(\bar{k}_4) \\
        &\hspace{3cm}+ \bar{\rho}^{(8)}_{\bbS}[34][12]  \cdot \left[  \eP^{{}_{(2)}}(\bar{k}_1)- \eP^{{}_{(1)}}(\bar{k}_1)\right]  \otimes \fF^{{}_{(1)}}(\bar{k}_2) \otimes \eP^{{}_{(2)}}(\bar{k}_3) \otimes \fF^{{}_{(1)}}(\bar{k}_4) \\
        &\hspace{3cm}  - \bar{\rho}^{(8)}_{\bbS}[12][34]  \cdot \left[  \eP^{{}_{(2)}}(\bar{k}_1)- \eP^{{}_{(1)}}(\bar{k}_1)\right]  \otimes \fF^{{}_{(1)}}(\bar{k}_2) \otimes \eP^{{}_{(1)}}(\bar{k}_3) \otimes \fF^{{}_{(1)}}(\bar{k}_4)
        \Big\}  \ ,
    \end{split}
\end{equation}
where we have defined
\begin{equation}
    \begin{aligned}
        \bar{\rho}^{(1)}_{\bbS}[43][21] &\equiv  \  
        n^{\fd}_{k^{0}_3+k^0_4} \langle \Gin \Sbout   \bbSA  \Gin \Sout \rangle  \ ,
        &&\bar{\rho}^{(1)}_{\bbS}[21][43] = - e^{-\beta k^0_2 -\beta k^0_1} \bar{\rho}^{(1)}_{\bbS}[43][21] \ ,\\
       \bar{\rho}^{(2)}_{\bbS}[21][43] &\equiv  \  
        (1-n^{\fd}_{k^{0}_3+k^0_4})\langle \Gin \Sbout   \bbSR  \Gin \Sout \rangle  \ ,  
        &&\bar{\rho}^{(2)}_{\bbS}[43][21] = - e^{-\beta k^0_4 -\beta k^0_3} \ \bar{\rho}^{(2)}_{\bbS}[21][43] \ ,\\
        \bar{\rho}^{(3)}_{\bbS}[43][12] &\equiv  \  
        n^{\fd}_{k^{0}_3+k^0_4}\langle \Gout \Sbin   \bbSA  \Gin \Sout \rangle  \ ,
        &&\bar{\rho}^{(3)}_{\bbS}[12][43] = - e^{-\beta k^0_2 -\beta k^0_1} \bar{\rho}^{(3)}_{\bbS}[43][12] \ ,\\
        \bar{\rho}^{(4)}_{\bbS}[12][43] &\equiv  \  
        (1-n^{\fd}_{k^{0}_3+k^0_4} ) \langle \Gout \Sbin   \bbSR  \Gin \Sout \rangle  \ , 
        &&\bar{\rho}^{(4)}_{\bbS}[43][12] = - e^{-\beta k^0_4 -\beta k^0_3} \bar{\rho}^{(4)}_{\bbS}[12][43] \ , \\
        \bar{\rho}^{(5)}_{\bbS}[34][21] &\equiv  \  
        n^{\fd}_{k^{0}_3+k^0_4}\langle \Gin \Sbout   \bbSA  \Gout \Sin \rangle  \ ,
        &&\bar{\rho}^{(5)}_{\bbS}[21][34] = - e^{-\beta k^0_2 -\beta k^0_1} \bar{\rho}^{(5)}_{\bbS}[34][21] \ ,\\
        \bar{\rho}^{(6)}_{\bbS}[21][34] &\equiv  \  
        (1-n^{\fd}_{k^{0}_3+k^0_4} ) \langle \Gin \Sbout   \bbSR  \Gout \Sin \rangle  \ , 
        &&\bar{\rho}^{(6)}_{\bbS}[34][21] = - e^{-\beta k^0_4 -\beta k^0_3} \bar{\rho}^{(6)}_{\bbS}[21][34] \ , \\
        \bar{\rho}^{(7)}_{\bbS}[34][12] &\equiv  \  
        n^{\fd}_{k^{0}_3+k^0_4}\langle \Gout \Sbin   \bbSA  \Gout \Sin \rangle  \ ,
        &&\bar{\rho}^{(7)}_{\bbS}[12][34] = - e^{-\beta k^0_2 -\beta k^0_1} \bar{\rho}^{(7)}_{\bbS}[34][12] \ , \\
        \bar{\rho}^{(8)}_{\bbS}[12][34] &\equiv  \  
        (1-n^{\fd}_{k^{0}_3+k^0_4} ) \langle \Gout \Sbin   \bbSR  \Gout \Sin \rangle  \ , 
        &&\bar{\rho}^{(8)}_{\bbS}[34][12] = - e^{-\beta k^0_4 -\beta k^0_3} \bar{\rho}^{(8)}_{\bbS}[12][34] \ .
    \end{aligned} 
\end{equation}

\bibliography{FermionEFTReferences}

\end{document}